\documentclass{iopjournal}
\usepackage{graphicx}
\usepackage{dcolumn}
\usepackage{latexsym}
\usepackage{hyperref}
\usepackage{amsmath, amsthm, amssymb}
\usepackage{relsize}
\usepackage{epsfig}
\usepackage{bm}
\usepackage[dvipsnames]{xcolor}
\usepackage{placeins}
\usepackage[normalem]{ulem}
\usepackage{diagbox}
\usepackage[normalem]{ulem}
\usepackage{tikz}

\begin{document}

\articletype{Article type} 

\title{Spatiotemporal Chaos and Defect Proliferation in Polar-Apolar Active Mixture}

\author{Partha Sarathi Mondal$^{a}$\orcid{https://orcid.org/0009-0009-1414-288X}, Tamás Vicsek$^{b}$\orcid{https://orcid.org/0000-0003-1431-2884} and Shradha Mishra$^{*,a}$\orcid{https://orcid.org/0000-0002-6069-9031}}

\affil{$^a$ Department of physics, Indian Institute of Technology (BHU) Varanasi, India 221005}\\
\affil{$^b$ Department of Biological Physics, Institute of Physics, Eötvös University, Budapest, H-1117, Hungary}\\

\vspace{1em}

\noindent

\noindent
\email{$^1$parthasarathimondal.rs.phy21@itbhu.ac.in, $^2$vicsek@hal.elte.hu, $^3$smishra.phy@iitbhu.ac.in}


\begin{abstract}
Chaotic transitions in inertial fluids typically proceed through a direct energy cascade from large to small scales. In contrast, active systems—composed of self-propelled units—inject energy at microscopic scales and therefore exhibit an inverse cascade, giving rise to distinctly unconventional flow patterns. Here, we investigate an active mixture consisting of both apolar and polar self-driven components, a setting expected to display richer behaviours than those found in living liquid crystal (LLC) systems, where the apolar constituent is passive. Using numerical solutions of the corresponding hydrodynamic equations, we uncover a variety of complex dynamical states. Our results reveal a non-monotonic response of the apolar species to changes in the density and activity of the polar component. In an intermediate regime—reminiscent of LLC-induced disorder—the system develops a dynamically disordered phase characterised by high-density, chaotically evolving band-like structures and by the continual creation and annihilation of ±1/2 topological defects. We show that this regime exhibits spatiotemporal chaos, which we quantify through two complementary measures: the spectral properties of density fluctuations and the maximal Lyapunov exponent. Together, these findings broaden the understanding of complex transitions in active matter and suggest potential experimental realisations in bacterial suspensions or synthetic microswimmer assemblies.
\end{abstract}

\section{Introduction\label{secI}}
Active matter systems comprise autonomous agents, called self-propelled particles (SPPs), that harness internal energy to generate spontaneous mechanical motion, driving the system out of equilibrium \cite{vicsek1995,bechinger2016active,bar2020self,ganguly2013aspects,vicsek2012collective}. The growing interest in active matter stems from its presence across a wide range of length scales \cite{alberts2002self,kolomeisky2007molecular,niu2012bacterial,shapiro1995significances,jolles2017consistent,moreno2020search,gueron1996dynamics,wang2019collective}. One of the simplest experimental realisations of active matter is a suspension of microswimmers, which serves as a valuable tool for understanding the fundamental principles governing the emergence of collective motion. Additionally, such systems provide a means to explore phase behaviour by precisely tuning key control parameters, including microswimmer concentration, self-propulsion velocity, and fluid viscosity \cite{sokolov2012physical,dunkel2013fluid}.\\
While microswimmer suspensions are typically studied in isotropic Newtonian fluids, the medium's anisotropy can significantly influence their individual and collective dynamics. These systems have attracted considerable interest due to their potential applications across several domains. A prominent example of such systems is the suspension of bacteria in Liquid Crystal (LC) solution, where bacterial activity perturbs LC orientation, triggering instabilities that drive a highly dynamic steady state — referred to as Living Liquid Crystal (LLC) \cite{gruler1995migrating,zhou2014living,turiv2020polar,kumar2013motility}. Microswimmers tend to align with the LC director, allowing control over their motion via surface anchoring. Increasing microswimmer concentration induces a transition to a chaotic state akin to active turbulence\cite{doostmohammadi2017onset,de2025hidden,shankar2022topological,alert2022active} in Active Nematics (AN), the active counterpart of LC. \\
Significant theoretical efforts have been made to deepen the understanding of the underlying physics governing the dynamics of these systems. One of the earliest theoretical contributions to modelling LLCs was made by Genkin et al. \cite{genkin2017topological}, who proposed a Coarse Grained Model based on the following assumptions: (i) the microswimmer concentration remains low, and they align with the local LC director, (ii) microswimmer motion does not directly influence the LC orientation. Within this regime, the model successfully reproduces experimentally observed behaviours. However, at higher bacterial concentrations, direct interactions with LC molecules become significant, driving the system into a distinct steady state characterised by symbiotic dynamics as demonstrated in Ref.\cite{vats2023symbiotic}.\\
The LLC framework can be extended to study the behaviour of microswimmers in anisotropic active media, such as AN.  
 Here we would like to emphasise one key difference from AN in contrast to LLC; in AN, particles themselves are self-propelled with head-tail symmetry, hence they are intrinsically active, whereas in LLC, the active nature of the liquid crystal particles comes due to the presence of foreign polar active particles. AN is prevalent in a wide range of biological and synthetic systems, including cytoskeleton filaments, cellular suspensions, biological tissues, etc \cite{kemkemer2000elastic,balasubramaniam2022active,saw2018biological}. Given the ubiquity of these systems in nature, studying suspensions of microswimmers within AN could provide critical insights into their behaviour in biologically and technologically relevant environments. Recent studies \cite{sampat2021polar,mondal2024dynamical,sharma2025motile} have shown that in such systems, the dynamics of one species is strongly affected by that of the other species. Motivated by this, we propose a coarse-grained model for microswimmers suspended in AN. Recent studies of similar systems using a microscopic approach have demonstrated that tuning the density and activity of microswimmers induces distinct phases in AN, including a dynamical disordered state with transient swirl-like structures \cite{sampat2021polar,mondal2024dynamical}. While previous studies have primarily examined the structural and statistical properties of AN, the dynamical aspects of the system demand further exploration—particularly in the dynamic disordered state. A key distinguishing feature of our model, compared with previous microscopic studies, is the nature of interspecies interactions.\\

In this study, we adopt a coarse-grained approach to model microswimmer suspensions in 2D AN. Our model does not account for hydrodynamic interactions between the particles, thereby violating momentum conservation. The possible experimental systems are the granular systems, where the particle dynamics is through partially inelastic collisions, or the particles typically move in direct contact with a solid substrate, which acts as a momentum sink  \cite{marchetti2013hydrodynamics,narayan2007long,kumar2019trapping}.  Even in systems where motion occurs close to the substrate in a viscous fluid bath, hydrodynamic interactions can be strongly suppressed when the friction coefficient is sufficiently large relative to the fluid viscosity \cite{marchetti2013hydrodynamics,alert2022active,doostmohammadi2019coherent,nejad2021memory}. Consequently, the essential physics of such systems can be captured by models that neglect the surrounding fluid. Accordingly, we model the active polar–apolar mixture in this limit.\\
The dynamics of the microswimmers are governed by a Toner-Tu-like model, where the mean density and self-propulsion speed serve as the primary control parameters. The AN is modelled using the coupled advection-diffusion equation for the density and nematic tensor order parameter field. Similar to LLC systems, the microswimmers are present in the minority. While LLC models assume a high-density regime for the LC -- allowing density fluctuations to be neglected -- our study considers AN in a low-density regime, where density fluctuations play a crucial role. These fluctuations establish a positive feedback loop between the density and orientation fluctuations of AN, leading to rich emergent behaviour.\\
The main findings of our study are as follows: (i) AN shows a reentrant behaviour with respect to the microswimmer density, with an inhomogeneous state emerging at intermediate densities; (ii) In the inhomogeneous regime, AN exhibits Giant Number Fluctuations and forms high-density nematically ordered bands in a low density disordered background; (iii) At higher microswimmer activity, the inhomogeneous regime transitions into a \emph{dynamic steady state}, characterised by persistent modulation of the bands and spontaneous formation and annihilation of $\pm\frac{1}{2}$ topological defects; (iv) The dynamics of AN in the \emph{dynamic steady state} exhibit spatiotemporal chaos, which is quantified by a positive value of the maximal Lyapunov Exponent.\\
The rest of the paper is structured as follows: Sec.\ref{sec:mod} describes the model and simulation details, and Sec. \ ref {sec:res} presents the results. Finally, Sec.\ref{sec:dis} summarises the study and discusses potential directions for future research.

\section{Model}\label{sec:mod}
We consider a suspension of microswimmers in an active nematic (or apolar species) on a $2D$ substrate wherein the microswimmers are modelled as polar species, and the system is referred to as a polar-apolar mixture hereafter. Each species is represented by its own set of slow variables: the density and order parameter. The density field of the polar and apolar species is denoted by $\rho_p(\boldsymbol{r})$ and $\rho_n(\boldsymbol{r})$, respectively. The local orientation of the polar species is given by $\hat{\boldsymbol{n}}_p(\boldsymbol{r})=(\cos\theta_p(\boldsymbol{r}),\sin\theta_p(\boldsymbol{r}))$, where $\theta_p$ denotes the orientation. The local order parameter for the polar species is given by the local polarisation vector, $\boldsymbol{P}(\boldsymbol{r}) = P_0(\boldsymbol{r}) \hat{\boldsymbol{n}}_p(\boldsymbol{r})$, where $P_0(\boldsymbol{r})$ is the magnitude of the local order parameter, and is also a function of $\boldsymbol{r}$. The unit vector $\hat{\boldsymbol{n}}_p(\boldsymbol{r}) = \frac{\boldsymbol{P}(\boldsymbol{r})}{\lvert \boldsymbol{P}(\boldsymbol{r}) \rvert} = \frac{\boldsymbol{P}(\boldsymbol{r})}{P_0(\boldsymbol{r})}$ denotes the local direction of ordering. For apolar species, the local director is given by $\hat{\boldsymbol{n}}_n(\boldsymbol{r})=(\cos\theta_n(\boldsymbol{r}),\sin\theta_n(\boldsymbol{r}))$. The local order parameter for the apolar species is represented by the nematic tensorial order parameter $\boldsymbol{Q}(\boldsymbol{r})$ owing to the head-tail symmetry of the apolar particles (i.e. $\hat{\boldsymbol{n}}_n(\boldsymbol{r}) \to - \hat{\boldsymbol{n}}_n(\boldsymbol{r})$) \cite{de1993physics}.\\
In the mixture, the state of the system is described by $(\rho_p, \rho_n, \boldsymbol{P}, \boldsymbol{Q})$, in which the inter-species interactions couple the dynamics of the polar and apolar particles. Experimental studies on LLC systems have shown that microswimmers tend to align nematically with the local director \cite{mushenheim2014dynamic}. This effect can be captured, to the leading order, by the free energy term $-(\boldsymbol{Q}: \mathbb{P}$), where $\mathbb{P}=(\boldsymbol{P}\boldsymbol{P} - \frac{\mathbb{I}}{2}\vert\boldsymbol{P}|^2)$ and  $\mathbb{I}$ is an identity tensor. The form of this interaction term is motivated by analogous behaviour observed in magnetic dopant particles suspended in a nematic liquid crystal host \cite{brochard1970theory,lopatina2009theory,bisht2020tailored}. In such systems, the polarisation vector $\boldsymbol{P}$ of the dopant acts as a local field influencing the orientation of the liquid crystal molecules, effectively favouring alignment between $\boldsymbol{P}$ and the nematic director $\hat{\boldsymbol{n}}_n$. This results in an energetic contribution of the form $-(\hat{\boldsymbol{n}}_n \cdot \boldsymbol{P})^2$, which, when expressed in terms of the nematic order parameter tensor $\boldsymbol{Q}$, leads to the aforementioned form of the term \cite{lopatina2009theory}.\\
The evolution equations for the density and the symmetry-broken variable for both species are discussed below.\newline 
\noindent
\uline{\emph{Equation for density fields}}: The evolution of the density fields is governed by the continuity equation,
\begin{equation*}
    \partial_t \rho_{\alpha} + \boldsymbol{\nabla}.\boldsymbol{J}_{\alpha} = 0
\end{equation*}
where $\alpha = (p, n)$ and the current $\boldsymbol{J}_{\alpha}$ has two contributions: a diffusive term, which accounts for particle transport due to random motion, and an active term. \\
For the polar species, the active current represents directed transport of particles along the local ordering direction defined by $\boldsymbol{P}$. It has the form $\boldsymbol{J}_p = v_p\rho_p \boldsymbol{P}$, where $v_p$ is the self-propulsion speed of the polar particles. The equation for polar density, $\rho_p$, is given by,
\begin{equation}
    \partial_t \rho_p = \boldsymbol{\nabla}.\bigg(D_{\rho_p}\boldsymbol{\nabla}\rho_p - v_p \rho_p \boldsymbol{P}\bigg)
    \label{eq:pden}
\end{equation} 
where, $D_{\rho_p}$ is the diffusion coefficient.\\
For the apolar species, the active contribution to $\boldsymbol{J}_n$ comes from the curvature induced density current $\boldsymbol{J}_a \sim -\rho_n\boldsymbol{\nabla}.\boldsymbol{Q}$, introduced by Simha et al. \cite{aditi2002hydrodynamic,ramaswamy2003active} and later also derived in ref.\cite{bertin2013mesoscopic}. The equation for apolar density, $\rho_n$, is given by, 
\begin{equation}
  \partial_t \rho_n = D_{\rho_n} \nabla^2 \rho_n + a_1 \boldsymbol{\nabla}.(\boldsymbol{Q}.\boldsymbol{\nabla}\rho_n) + a_3 \boldsymbol{\nabla}.(\rho_n \boldsymbol{\nabla}.\boldsymbol{Q})  
    \label{eq:nden}
\end{equation}
where $D_{\rho_n}$ and $a_1$ are the coefficients of isotropic and anisotropic diffusion, respectively. The coefficient $a_3$ represents the activity of the apolar species.\newline 
\noindent
\uline{\emph{Equation for order parameter fields}}: In an equilibrium system, the dynamics of the symmetry broken variable $\psi$ is governed by the gradient descent on the landscape described by the Landau-Gingzburg free energy functional \cite{goldenfeld2018lectures}, 
\begin{equation*}
    F_{\psi}\{{\boldsymbol{\psi}}(\boldsymbol{r})\} = \mathlarger{\int} \bigg[\frac{1}{2}\alpha_{\psi} \psi^2 + \frac{1}{4}\beta_{\psi} \psi^4  + k_{\psi} (\boldsymbol{\nabla}\psi)^2 \bigg] d^{2}\boldsymbol{r}
\end{equation*}
where, $\psi$ stands for the order parameter, $\psi = {\boldsymbol{P}}$ for the polar species, and $\boldsymbol{Q}$ for the apolar species. For polar species, $\psi^2 = \mathlarger{\sum}_i P_i P_i$, $\psi^4 = (\sum_i P_i P_i)(\sum_j P_j P_j)$, and $(\boldsymbol{\nabla}\psi)^2=\mathlarger{\sum}_{i,k} (\partial_k P_{i})(\partial_k P_{i})$. For apolar species, $\psi^2 = \mathlarger{\sum}_{i,j}Q_{ij}Q_{ij}$, $\psi^4 = (\sum_{i,j}Q_{ij}Q_{ij}) (\sum_{k,l}Q_{kl}Q_{kl})$, and $(\boldsymbol{\nabla}\psi)^2=\mathlarger{\sum}_{i,j,k}(\partial_k Q_{ij})(\partial_k Q_{ij})$. Where, $i,j,k \equiv x,y$ in $d=2$.\\
The first two terms control the mean field order-disorder transition with $\beta_{\psi} > 0$ and $\alpha_{\psi} = \bigg(\rho_{\psi c}-\rho_{\psi}\bigg)$, $\rho_{\psi c} = \rho_{pc}$, $\rho_{nc}$ is the mean field critical density for the polar and apolar species, respectively. For $\alpha_{\psi}>0$, the steady state is disordered, and for $\alpha_{\psi}<0$, the steady state is a homogeneous ordered state, which is ferromagnetically ordered for polar species and nematically ordered for apolar species. The third term penalises the distortions in the $\boldsymbol{P}$ or $\boldsymbol{Q}$ fields with $k_{p,n}>0$. The form of the distortion term is obtained from the Frank free energy functional under the unified elastic constant approximation \cite{de1993physics}.\\
Considering the effect of coupling, the total free energy functional is given by
\begin{equation*}
    F_{tot} = F_{\boldsymbol{P}} + F_{\boldsymbol{Q}} - \gamma \int(\boldsymbol{Q}:\mathbb{P})d^2\boldsymbol{r}
\end{equation*}
where $\gamma$ is the strength of coupling and $\mathbb{P}=(\boldsymbol{P}\boldsymbol{P} - \frac{\mathbb{I}}{2}\vert\boldsymbol{P}|^2)$.\\
The equation for the nematic tensor order parameter, $\boldsymbol{Q} \equiv (Q_{xx},Q_{xy})$, for the apolar species is given by,
\begin{equation}
\partial_t Q_{ij}
= -\Gamma_Q \frac{\partial F_Q}{\partial \boldsymbol{Q}}
+ \frac{D_3}{2\rho_{n0}} \left(\partial_i \partial_j - \frac{1}{2}\delta_{ij}\nabla^2\right)\rho_n
+ \gamma \left(P_i P_j - \frac{1}{2}\delta_{ij} P_k P_k\right)
\label{eq:nop}
\end{equation}

In the above equation, the first term denotes the relaxational dynamics of the nematic order parameter, where $\Gamma_Q$ decides the timescale of relaxation. The second term describes the change in components of $\boldsymbol{Q}$ due to inhomogeneities in the density field $\rho_n$, and the third term represents the contribution of coupling.\\
The equation for the polarization field $\boldsymbol{P} \equiv (P_x,P_y)$ for the polar species is given by-

\begin{eqnarray}
\partial_t \boldsymbol{P}
+ \lambda_1 (\boldsymbol{P}\!\cdot\!\boldsymbol{\nabla})\boldsymbol{P}
+ \lambda_2 (\boldsymbol{\nabla}\!\cdot\!\boldsymbol{P})\boldsymbol{P}
+ \lambda_3 \boldsymbol{\nabla}(|\boldsymbol{P}|^2)
&=& \Gamma_p \Big[(-\alpha_p - \beta_p |\boldsymbol{P}|^2)\boldsymbol{P}
+ k_p \nabla^2 \boldsymbol{P}\Big] - \frac{\sigma_1}{2\rho_{p0}} \boldsymbol{\nabla}\rho_p\nonumber\\
&&+ D_B \boldsymbol{\nabla}(\boldsymbol{\nabla}\!\cdot\!\boldsymbol{P})
+ D_2 (\boldsymbol{P}\!\cdot\!\boldsymbol{\nabla})^2 \boldsymbol{P}
+ 2\gamma (\boldsymbol{Q}\!\cdot\!\boldsymbol{P})
\label{eq:pop}
\end{eqnarray}

For polar species, $\boldsymbol{P}$ serves as both the orientational order parameter and a source of directional bias, leading to the advection of polarisation on local scales. In the absence of activity, the dynamics of $\boldsymbol{P}$ is governed by relaxational dynamics on a free energy landscape described by the free energy functional $F_{\boldsymbol{P}}$, $-\Gamma_p \frac{\delta F_p}{\delta\boldsymbol{P}}$, with the advection term $\lambda_1(\boldsymbol{P}.\boldsymbol{\nabla})\boldsymbol{P}$. The terms inside the square bracket on the right-hand side describe the free energy dominated relaxation of $\boldsymbol{P}$ within the time scale $\frac{1}{\Gamma_p}$. The presence of activity breaks Galilean invariance, allowing the coefficient of the advection term, $\lambda_1$, to differ from unity. Additionally, other terms of the same order in $\boldsymbol{P}$ and $\boldsymbol{\nabla}$ as the advection term, consistent with the symmetry and conservation laws, can be included in the dynamical equation. Hence, the terms with coefficients $\lambda_2$ and $\lambda_3$ are included. The term with coefficients $\sigma_1$ and $D_2$ represents the contribution of the pressure set up by the density gradient and anisotropic diffusion along the direction of $\boldsymbol{P}$, respectively. The $D_B$ term is of the same order in $\boldsymbol{\nabla}$ and $\boldsymbol{P}$ as the distortion term with coefficient $k_p$, and is allowed due to the breakdown of Galilean invariance. In the context of polar flock, the $D_B$ and $k_p$ terms are analogous to the bulk and shear viscosity terms \cite{toner2022walking} in a compressible fluid. The last term in Eq.\ref{eq:pop} represents the effect of coupling on the dynamics of the $\boldsymbol{P}$-field. \newline

\noindent
\uline{\emph{Parameter and simulation details}}: In our model, the critical densities of the polar and the apolar species are set to $\rho_{pc}=\rho_{nc}=0.50$. The mean density, $\rho_{n0}$, and activity of the apolar species, $a_3$, are kept fixed at $\rho_{n0} = 0.75$ and $a_3=0.60$, respectively. The mean density, $\rho_{p0}$, and the self propulsion speed, $v_p$, of the polar species is varied in the range $\rho_{p0}\in[0.0001,0.20]$, and $v_p\in[0.001,0.70]$. The choice of phenomenological parameters in Eqs.(\ref{eq:pden}-\ref{eq:pop}) is discussed in Appendix \ref{app:A}. We fixed the other parameters and varied only the density and activity of polar species. In both pure polar and pure nematic active systems, it is well established that the instability leading to pattern formation is primarily controlled by the density and activity \cite{mishra2010fluctuations,putzig2014phase,marchetti2013hydrodynamics,alert2022active}. The remaining parameters mainly influence quantitative aspects of the instability threshold and the resulting structures, but do not significantly modify the qualitative behaviour \cite{mishra2010fluctuations,putzig2014phase}. The intrinsic time and length scales are obtained from relaxation time  $t_0 = \Gamma_Q^{-1}$ and $l_0 = \sqrt{D_{\rho_n} \Gamma_Q}$, and in the present study, both of them are chosen as one. All other time and length scales are made dimensionless using $t_{0}$ and $l_{0}$.\\

\begin{figure*}[hbt]
    \centering
    \includegraphics[width=0.995\linewidth]{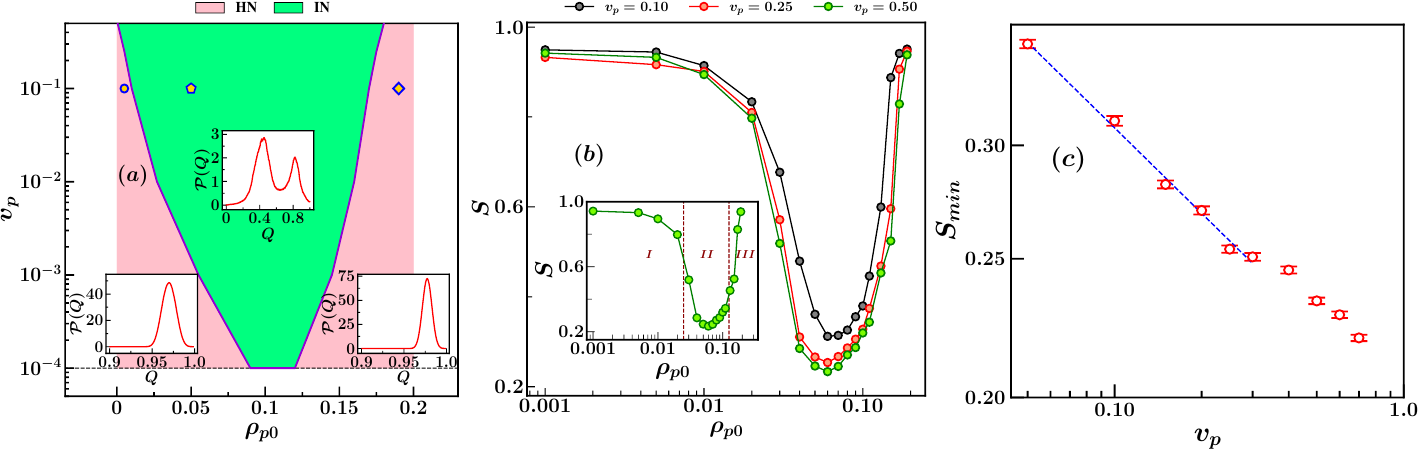}
    \includegraphics[width=0.99\linewidth]{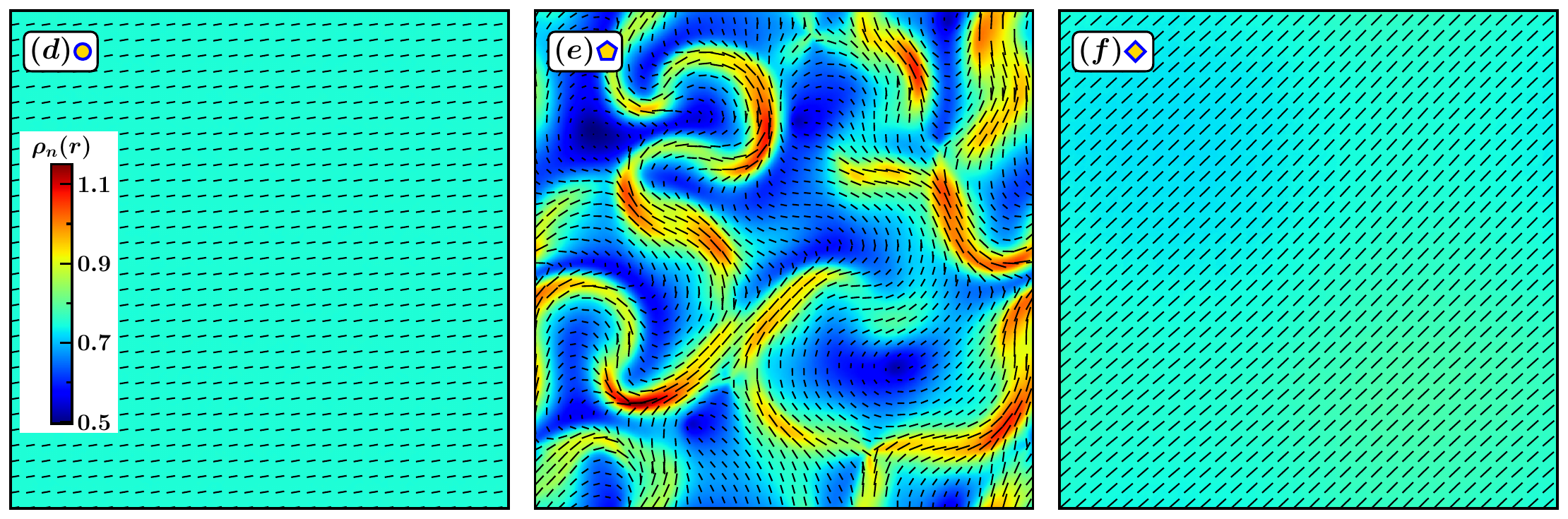}
    \caption{Global characteristics of active nematics containing an active polar component. Panel (a) shows the phase diagram of the system based on the steady-state characteristics of the active nematic. In the homogeneous regime, both $\rho_n$ and $\boldsymbol{Q}$ are homogeneous, whereas in the IN-regime $\rho_n$ and $\boldsymbol{Q}$ exhibit significant spatial fluctuations. The insets show the probability density function of $Q = |\boldsymbol{Q}|$, $P(Q)$. In the homogeneous regime, $P(Q)$ shows a single peak, whereas in the IN-regime, $P(Q)$ shows two peaks showing the coexistence of high and low ordered regions. The panel (b) shows the variation of nematic scalar order parameter, $S$, with polar density, $\rho_{p0}$, for different values of $v_p$. The entire range of $\rho_{p0}$ is partitioned into distinct states according to the magnitude of $S$, as illustrated in the inset. The panel (c) showcases the plot of $S_{min}$ $vs.$ $v_p$ on a log–log scale, with the dashed line representing a power-law fit. The panels (d-f) show the snapshots of the density and orientation field of apolar species in different phases : (d) Phase-I, (e) Phase-II, (f) Phase-III. The heat map represents the density field, while the lines indicate the local nematic director $\hat{n}_n=(\cos(\theta_n),\sin(\theta_n))$, with their length proportional to the local nematic order $\vert \boldsymbol{Q}\vert$. The snapshots show the full simulation box. The PDF $P(Q)$ and $P(\delta \rho_n)$ corresponding to the snapshots (d-f) are shown in figure \ref{fig:histdennop}. Parameters: System size, $L = 512$.}
    \label{fig:pd_nop}
\end{figure*}

\begin{figure}[hbt]
    \centering
    \includegraphics[width=0.99\linewidth]{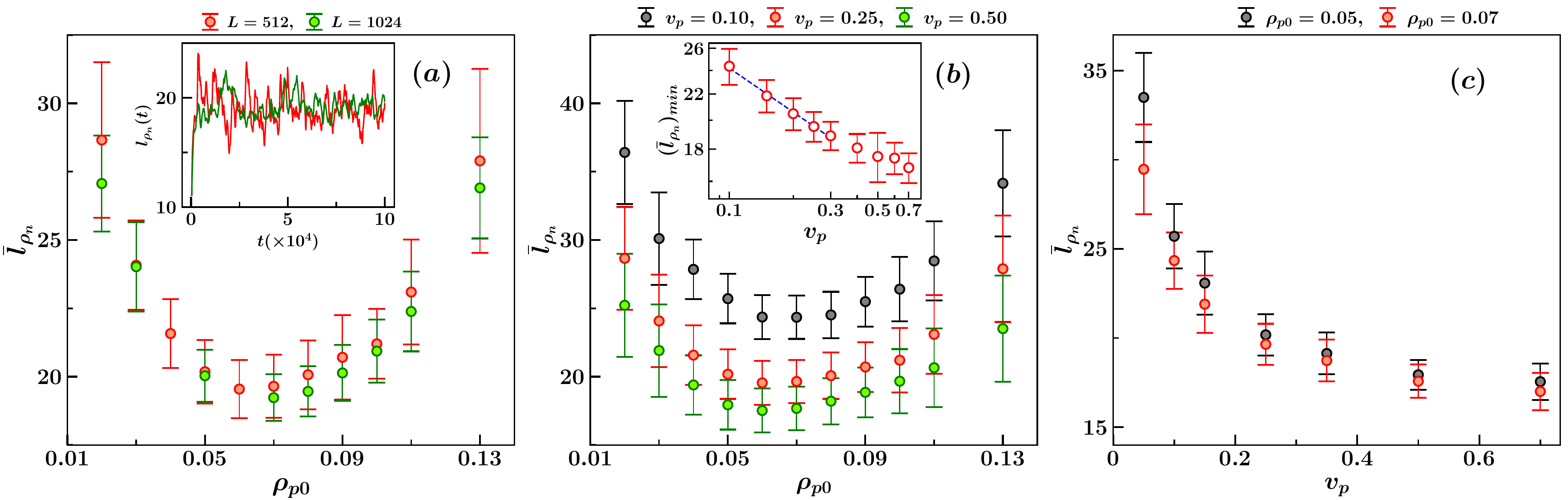}
    \caption{Variation of structural properties of the bands with changing control parameters in the IN regime. Panel (a) shows the variation of averaged correlation length, $<l_{\rho_n}>$, with polar density, $\rho_{p0}$, in Phase-II for two different system sizes $L =512$, $\&$ $1024$ for $v_p = 0.25$. The inset shows the time series of $l_{\rho_n}(t)$ for $v_p = 0.25$ and $\rho_p = 0.07$; Panel (b) presents $\overline{l}_{\rho_n}$ $vs.$ $\rho_{p0}$ plot for different values of $v_p$. The inset of subplot (b) depicts the plot of $\langle l_{\rho_n} \rangle_{min}$ $vs.$ $v_p$. The dashed line represents the power law behaviour; Panel (c) depicts the $\overline{l}_{\rho_n}$ $vs.$ $v_p$ plot for two different values of $\rho_{p0}$ in Phase-II. Parameters: System size, $L = 512$ for subplots (b) and (c). The rest of the parameters are the same as in figure \ref{fig:pd_nop}.}
    \label{fig:nem_corr_len_den}
\end{figure}

We simulate Eq.(\ref{eq:pden}-\ref{eq:pop}) on a square lattice of size $L$ with periodic boundary conditions in both directions. For numerical integration, we used the Euler method with space and time grid sizes $\Delta x$ and $\Delta t$, respectively, satisfying the stability criterion $\frac{\Delta t}{(\Delta x)^2} < \frac{1}{2}$. In our simulation, we use $\Delta x = 1.0$ and $\Delta t = 0.1$. We have checked that for a small variation in $\Delta t$, the results do not change. One simulation step consisted of updating the state of the system (i.e., $\{\rho_p, \rho_n, \boldsymbol{P}, \boldsymbol{Q}\}$ fields simultaneously) at each lattice point once. The data shown below are obtained for system sizes $L = 512$ and $1024$, unless stated otherwise. We start with a homogeneous and isotropic distribution of $(\rho_p, \rho_n, \boldsymbol{P}, \boldsymbol{Q})$. The system is simulated for $12 \times 10^5$ and  $50 \times 10^5$ time steps for $L = 512$, and $L = 1024$, respectively. In either case, the first $6 \times 10^5$ time steps are allocated to allow the system to reach the corresponding non-equilibrium steady state (see Appendix \ref{app:A}). Observables are calculated in the remaining time steps. For improved statistical accuracy, the data presented below (figures \ref{fig:pd_nop}, \ref{fig:nem_corr_len_den}, \ref{fig:auto_corr_fun}, \ref{fig:nem_defs}) are averaged over $50$ independent realisations.

\section{Results}\label{sec:res}

\begin{figure}[hbt]
    \centering
    \includegraphics[width=0.99\linewidth]{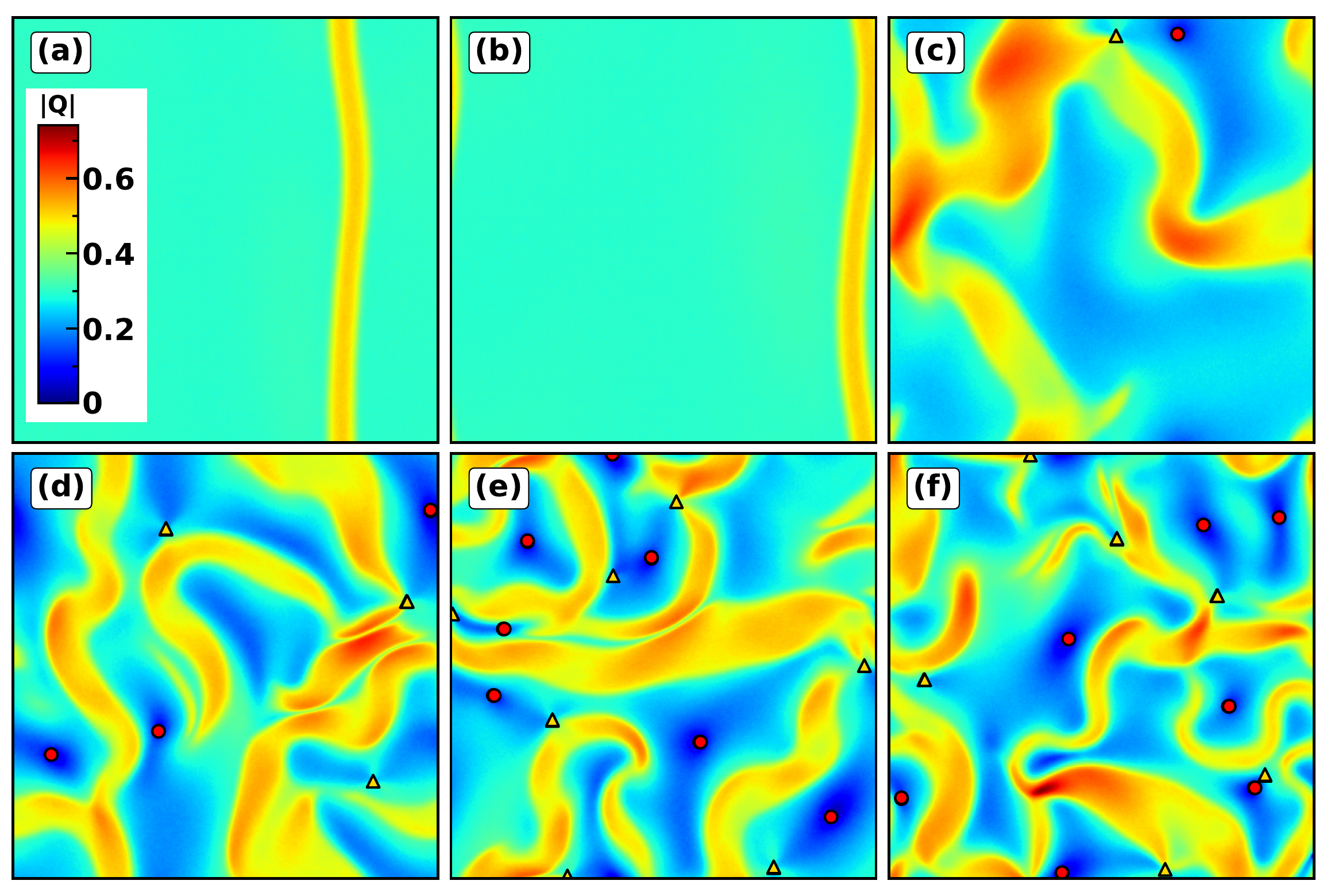}
    \caption{Snapshots of the magnitude of $\boldsymbol{Q}$-field (i.e. $Q$) for the apolar species for $\rho_{p0} = 0.09$ in Phase-II for different values of polar activity : (a) $v_p = 0.001$, (b) $v_p = 0.005$, (c) $v_p = 0.05$, (d) $v_p = 0.10$, (e) $v_p = 0.20$, and (f) $v_p = 0.50$. The heatmap depicts the magnitude of the nematic order parameter field $Q = |\boldsymbol{Q}|$ for active nematics. The snapshots show the full simulation box. In (c-f), the $\pm \frac{1}{2}$ defects are marked by circles and triangles, respectively. Parameters: System size, $L = 512$. The rest of the parameters are the same as in figure \ref{fig:pd_nop}.}
    \label{fig:phase2_diffact}
\end{figure}

\subsection{Phase Diagram}\label{sec:pd}
The phase diagram of the system in the $(\rho_{p0}, v_p)$ plane is shown in figure \ref{fig:pd_nop}(a), reveals three distinct regimes of the characteristics of the apolar species, referred to as Phase-I, II, and III respectively, as the mean density of the polar species, $\rho_{p0}$, is varied. Phases I and III are homogeneous nematically ordered states (HN). In contrast, Phase II is inhomogeneous (IN), wherein high-density nematically ordered domain(s), referred to as band(s), coexist with a low-density disordered background. In the IN-regime, for lower values of the self-propulsion speed, the apolar species forms a single, large band structure that extends along the length of the domain, with a width significantly smaller than its length. Inside the band, the local nematic directors are predominantly aligned along the long axis. In contrast, at higher values of $v_p$ ($v_p \gtrsim 0.05$), the system displays multiple smaller, narrow, elongated structures in place of the single extended band. The distinct spatiotemporal patterns associated with each phase are illustrated through snapshots in figure \ref{fig:pd_nop}(d-f), where the colour map and the lines are representative of the density field, $\rho_n$, and the local nematic director of the apolar species, respectively. For the polar species, the $\boldsymbol{P}$-field is aligned inside the band, while outside the band, its magnitude is $\approx 0$, indicating that the polar species is disordered outside the band. The spatial configuration of the polar species in different phases is shown in figure \ref{fig:psn_diff}.\\
To quantify the local nematic ordering of the apolar species, we calculated the magnitude of the local nematic order parameter, denoted by $Q(\boldsymbol{r})$, and orientation of local nematic director, $\theta_n(\boldsymbol{r})$, which are defined as --
\begin{equation}
    Q = Q(\boldsymbol{r}) = \sqrt{Q_{xx}^2(\boldsymbol{r})^2 + Q_{xy}^2(\boldsymbol{r})}
    \label{eq:lnop}
\end{equation}
 and 
 \begin{equation}
    \theta_n = \theta_n(\boldsymbol{r}) = \frac{1}{2}\arctan\bigg(\frac{Q_{xy}(\boldsymbol{r})}{Q_{xx}(\boldsymbol{r})}\bigg)
    \label{eq:ltheta}
\end{equation}
To emphasise the local inhomogeneity in nematic ordering, the probability density function (PDF) of the local nematic order parameter $Q$, denoted by $\mathcal{P}(Q)$, in different Phases is shown in the insets of figure \ref{fig:pd_nop}(a). In Phase-I and III, the $\mathcal{P}(Q)$ exhibits a single peak, whereas in Phase-II $\mathcal{P}(Q)$ is bimodal. Similar behaviour is observed across different phases for the PDF of the local density fluctuations of the apolar species, $\delta \rho_n(\boldsymbol{r}) = \rho_n(\boldsymbol{r}) - \rho_{n0}$, denoted by $\mathcal{P}(\delta\rho_n)$. 
The details of the calculation of PDF are discussed in Appendix \ref{app:pdf}, and the characteristics of the density fluctuation in the IN-regime are discussed in Appendix \ref{app:denfluc}. Additionally, in figure \ref{fig:histdennop} we show the PDFs obtained in Phase II for $\Delta t = 0.025$. The resulting PDFs match almost exactly with that of $\Delta t=0.1$ used in our simulations, which implies that the statistical characteristics of Phase II are robust with respect to modest variations in $\Delta t$.\\
The reentrant response of the apolar species with respect to $\rho_{p0}$ can be quantified by measuring the global nematic ordering of the system characterised by the nematic scalar order parameter as,
\begin{equation}
     S =\Bigg<2\sqrt{\bigg(\overline{\cos^2\theta_n}  -\frac{1}{2}\bigg)^2 + \bigg(\overline{\sin \theta_n \cos\theta_n}\bigg)^2} \Bigg>
\end{equation}
where the $\overline{\cdots}$ denotes an average over all lattice points, and $\langle \dots \rangle$ represents an average over time in the steady state and multiple independent realisations. $S \approx 1$ and $S \approx 0$ imply globally nematically ordered and disordered states, respectively.\\
The variation of $S$ with $\rho_{p0}$ for different values of $v_p$ is shown in the main window of the figure \ref{fig:pd_nop}(b). The inset of figure \ref{fig:pd_nop}(b) shows different phases with respect to the value of $S$ marked by drawing vertical dashed lines. In Phase-I, $S\approx 1$ signifies global nematic ordering in the system. In Phase-II, the value of $S$ drops to a value $\approx 0.20$, which is a consequence of the fact that the direction of the nematic ordering in different bands is different. On further increasing $\rho_{p0}$, the homogeneous ordered state is recovered in Phase-III.\\
It is evident from figure \ref{fig:pd_nop}(b) that the location of the minima of $S$ does not depend on $v_p$, whereas the depth of the minima increases with an increase in $v_p$. The value of $S$ at the location of minima is denoted by $S_{min}$. The plot of $S_{min}$ \emph{vs.} $v_p$ is shown in figure \ref{fig:pd_nop}(c) in log-log scale, which shows a power-law decay up to $v_p \approx 0.30$, beyond which the decay deviates from power-law behaviour.\\

\begin{figure}[hbt]
    \centering
    \includegraphics[width=0.8\linewidth]{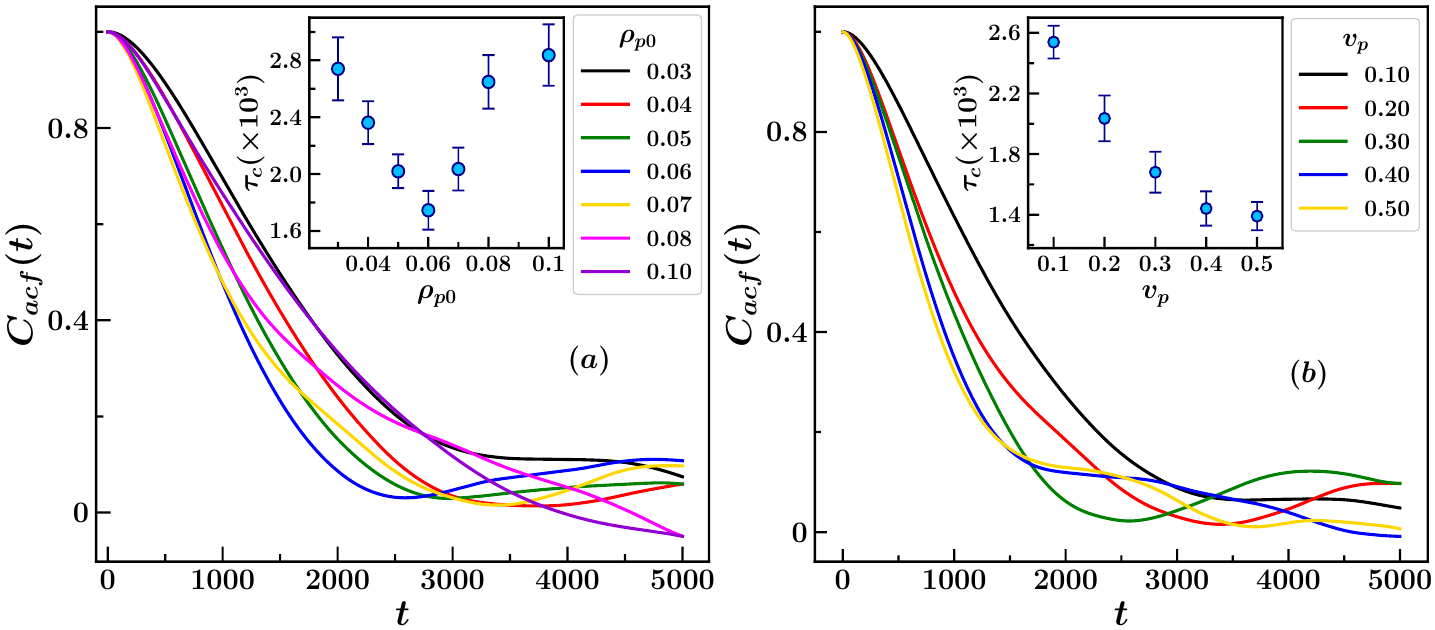}
    \caption{Plot of the autocorrelation function of the fluctuations in $l_{\rho_n}(t)$, $C_{acf}(t)$,\emph{vs.}time, $t$, for a different set of control parameters. Panel (a) shows $C_{acf}(t)$\emph{vs.}$t$ for different values of $\rho_{p0}$ for $v_p = 0.20$. The inset shows the plot of the correlation time, $\tau_c$, \emph{vs.} $\rho_{p0}$. Panel (b) shows $C_{acf}(t)$\emph{vs.}$t$ for different values of $v_p$ for $\rho_{p0} = 0.07$. The inset shows the plot of the correlation time, $\tau_c$, \emph{vs.} $v_p$. The error bars in the inset of (a) and (b) represent the standard deviation of $\tau_c$ calculated over independent realisations. System Size, $L=512$. The rest of the parameters are the same as in figure \ref{fig:pd_nop}.}
    \label{fig:auto_corr_fun}
\end{figure}

\subsection{Characterizing the bands in the inhomogeneous regime}\label{sec:inhden}
Spatial configurations of the apolar species for different values of $v_p$ in the inhomogeneous regime are shown in figure \ref{fig:phase2_diffact} through a series of snapshots of the magnitude of the local nematic order parameter. The snapshots clearly show that the bands are macroscopic in size. With an increase in $v_p$, a single band destabilises and forms multiple narrow bands as shown in figure \ref{fig:phase2_diffact}(a-c). On further increase of $v_p$, the smaller bands become progressively narrower, and the distortions in the nematic orientation field become more pronounced for $v_p \gtrsim 0.05$ as shown in figure \ref{fig:phase2_diffact}(c-f). Despite their macroscopic size, the bands undergo frequent modulations, including stretching, bending, merging, and splitting, in the system's steady state, which we call the \emph{dynamic steady state}. A visual description of the modulation of bands is presented through an animation in \hyperref[sm:mov1]{MOVIE-1} for different values of $v_p$, which shows that with an increase in $v_p$, the dynamics of the bands enhances, resulting in more frequent modulations.\\
The bands can be characterised by calculating the two-point correlation function of the density field, $\rho_n(\boldsymbol{r},t)$ and the order parameter field, $\boldsymbol{Q}(\boldsymbol{r},t)$ of the apolar species, defined as -
\begin{equation}
    C_{\rho_n}(r,t)=\langle\delta\rho_n(\boldsymbol{r}_0,t)\delta\rho_n(\boldsymbol{r}_0+\boldsymbol{r},t)\rangle
    \label{eq:dencorr}
\end{equation}
and 
\begin{equation}
    C_{Q}(r,t)=\langle\boldsymbol{Q}(\boldsymbol{r}_0,t):\boldsymbol{Q} (\boldsymbol{r}_0+\boldsymbol{r},t)\rangle
    \label{eq:opcorr}
\end{equation}
where, $\delta\rho_n(\boldsymbol{r},t) = \rho_n(\boldsymbol{r},t) - \rho_{n0}$ is the local density fluctuation of apolar species and $\langle \dots \rangle$ implies average over reference points $\boldsymbol{r}_0$, direction of $\boldsymbol{r}$ for a given $r=\vert \boldsymbol{r}\vert$ as well as different independent realisations. The correlation function is normalised with its value at $r=0$, and the corresponding correlation length is extracted as the distance at which the correlation function first decays to $0.5$. The correlation length is a characteristic length scale in the system and can be used to define the effective size of the structures (i.e., the bands) \cite{hohenberg1989chaotic}. The following analysis is presented in terms of the correlation length of the $\rho_n$-field, denoted by $l_{\rho_n}$; analogous results are obtained using the correlation length of the $\boldsymbol{Q}$-field, denoted by $l_{Q}$. The results are robust with respect to the specific cut-off criterion used to define the correlation length.\\
As the system evolves from a homogeneous state, $l_{\rho_n}(t)$ increases, reflecting the growth of high-density domains in the apolar species, before eventually saturating. In the steady state, $l_{\rho_n}(t)$ fluctuates around a mean value, indicating that the growth of the bands has plateaued as shown in the inset in figure \ref{fig:nem_corr_len_den}(a). The steady-state correlation length can be expressed as $\overline{l}_{\rho_n} = \langle l_{\rho_n}\rangle \pm \Delta l_{\rho_n}$ where, $\langle l_{\rho_n}\rangle$, $\Delta l_{\rho_n}$ are the mean and standard deviation of $l_{\rho_n}(t)$, respectively, obtained from the ensemble averaged PDF of $l_{\rho_n}(t)$. The mean $\langle l_{\rho_n}\rangle$ represents the effective size of bands in the steady state, while $\Delta l_{\rho_n}$ quantifies temporal modulations of the bands. To illustrate the same, simultaneous evolution of the $\rho_n(\boldsymbol{r})$ and $\boldsymbol{Q}(\boldsymbol{r})$ field of the apolar species and the corresponding correlation lengths, $l_{\rho_n}$ and $l_{Q}$, are shown through an animation in \hyperref[sm:mov2]{MOVIE-2}.\\
To ensure that the saturation of $l_{\rho_n}(t)$ is not a finite-size effect, we examined the behaviour for two different system sizes as shown in figure \ref{fig:nem_corr_len_den}(a). The inset of figure \ref{fig:nem_corr_len_den}(a) presents the time series of $l_{\rho_n}(t)$ $vs.$ $t$ for $L = 512$, $1024$. Saturation occurs within the same time range for both system sizes, and the mean steady-state value of $l_{\rho_n}(t)$ is approximately identical. Further, the plot of $\overline{l}_{\rho_n}$ $\emph{vs.}$ $\rho_{p0}$ for a fixed value of $v_p$ shows excellent agreement between the two system sizes as shown in the main panel of figure \ref{fig:nem_corr_len_den}(a). The consistency across system sizes confirms that the saturation of $l_{\rho_n}(t)$ is not a finite-size effect, and the mean value reflects the intrinsic length scale of the system, i.e., the characteristic length scale of the high-density nematically ordered regions of apolar species. The effective size of the bands is therefore independent of the system size and is determined solely by the choice of control parameters $(\rho_{p0}, v_p)$.\\
The main panel of figure \ref{fig:nem_corr_len_den}(b) presents the variation of the  $\overline{l}_{\rho_n}$ with $\rho_{p0}$ for various values of $v_p$. The associated error bars denote the standard deviation, $\Delta l_{\rho_n}$. A minimum in $\overline{l}_{\rho_n}$ is observed deep within the IN-regime, indicating that the effective size of band structures reaches a minimum for the corresponding set of control parameters. As $\rho_{p0}$ approaches the phase boundary on either side, the band broadens. The non-monotonic dependence of $\overline{l}_{\rho_n}$ on $\rho_{p0}$ is consistent with the reentrant behaviour of the global ordering and local structural characteristics of apolar species discussed so far. Moreover, the increase in the magnitude of the error bars near the phase boundaries reflects the temporal modulation of the bands over relatively larger length scales, resulting in larger fluctuations in the time series of $l_{\rho_n}(t)$.\\
The plot of $\overline{l}_{\rho_n}$ \emph{vs.}  $\rho_{p0}$ for different values of $v_p$ demonstrates that while the location of the minimum remains unchanged with increasing \(v_p\), the minima become more pronounced with an increase in $v_p$. The value of $\overline{l}_{\rho_n}$ at the minimum is denoted by $(\overline{l}_{\rho_n})_{\mathrm{min}}$. The inset of figure \ref{fig:nem_corr_len_den}(b) shows the variation of $(\overline{l}_{\rho_n})_{\mathrm{min}}$ with $v_p$ on a log-log scale. The observed decay follows a power-law behaviour up to approximately $v_p \approx 0.30$, beyond which deviations from the power-law behaviour become apparent. It is important to emphasise that the range of $v_p$ over which the power-law behaviour is observed coincides with the range shown in the range of power-law decay in the $S_{\mathrm{min}}$ \emph{vs.} $v_p$ plot shown in figure \ref{fig:pd_nop}(c). Although the limited range of $v_p$ (less than one order of magnitude) precludes a precise determination of the power-law exponent, a naive estimation yields similar values of exponents in both cases. The scaling behaviour is reminiscent of the turbulent states observed across a range of systems. In classical fluid turbulence, characteristic length scales, such as the typical vortex size, exhibit a power-law behaviour with their range depending on the Reynolds number, the relevant control parameter \cite{kolmogorov1995turbulence}. A similar trend is observed in active turbulence, where the intrinsic active length scale—governing the decay of velocity or vorticity correlations - scales as a power law with activity \cite{alert2022active,doostmohammadi2018active,giomi2015geometry}. These analogies suggest the connections of the spatiotemporally chaotic state to turbulence-like states reported in other systems.\\
Furthermore, the plot of $\overline{l}_{\rho_n}$ \emph{vs.} $v_p$, shown in figure \ref{fig:nem_corr_len_den}(c), indicates a decrease in effective size of bands with increasing $v_p$, consistent with the spatial patterns observed in the snapshots of figure \ref{fig:phase2_diffact}(d-f). The reduction in the magnitude of the error bars with increasing $v_p$ can be attributed to the modulation of the bands occurring on shorter length scales at higher activity.

\begin{figure*}[hbt]
    \centering
    \includegraphics[width=0.995\linewidth]{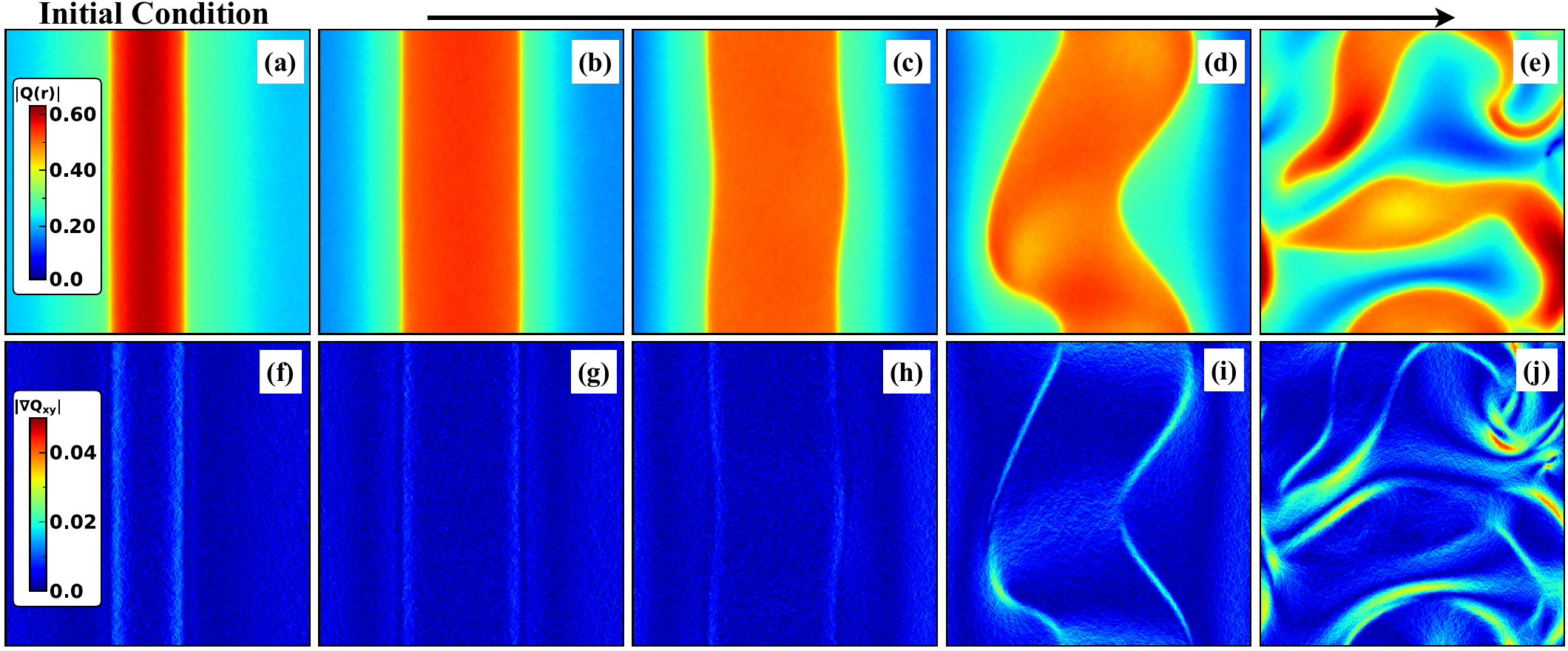}
    \caption{Mechanism of the formation of chaotic bands following an instantaneous change in the $v_p$. The system is prepared to have an initial condition that corresponds to the steady-state for $v_p=0.001$, as shown in panel (a), which shows the configuration of the nematic order parameter field $\vert \boldsymbol{Q}(\boldsymbol{r}) \vert$. The system is then instantaneously changed to $v_p =0.07$. The subsequent snapshots in the top row (b-e) show the temporal evolution of the magnitude of the nematic order parameter field  $\vert Q(\boldsymbol{r}) \vert$, and those in the bottom row (f-j) displays the local stress magnitude corresponding to the snapshot shown in the top row, $\sigma_{loc}(\boldsymbol{r})$. The snapshots show the full simulation box. Parameters : $L = 400$. The rest of the parameters are the same as in figure \ref{fig:pd_nop}.}
    \label{fig:band_formation}
\end{figure*}

\subsubsection{\uline{Dynamics of the bands} :} 
To check whether the temporal evolution of the structures in the system has any inherent time scale, we computed the autocorrelation function of the fluctuations in $l_{\rho_n}(t)$,  denoted by -

\begin{equation}
C_{acf}(t) = \langle \delta l_{\rho_n}(t_0+t)\delta l_{\rho_n}(t_0) \rangle  
\end{equation}
where, $\delta l_{\rho_n}(t) = l_{\rho_n}(t) - \overline{l}_{\rho_n}$ denotes the instantaneous deviation of $l_{\rho_n}(t)$ from its mean value $\overline{l}_{\rho_n}$, and $\langle \cdots \rangle$ indicate averaging over both the reference time $t_0$ and independent realisations. The autocorrelation function is normalised by its value at $t=0$. The correlation time $\tau_c$ is defined as the time at which the autocorrelation function drops to $\frac{1}{e}$ of its value at $t=0$, i.e, $C_{acf}(t = \tau_c)=C_{acf}(t = 0) e^{-1}$. Accordingly, the early time decay rate of the autocorrelation function is given by $r_c = 1 / \tau_c$. \\
The figure \ref{fig:auto_corr_fun} displays $C_{acf}(t)$ as a function of time, $t$, for various choices of control parameters. The main window in figure \ref{fig:auto_corr_fun}(a) shows the plot of $C_{acf}(t)$ \emph{vs.} $t$ for different $\rho_{p0}$, for $v_p = 0.20$ which reveals a non-monotonic dependence of the early-time decay rate of $C_{acf}(t)$ on $\rho_{p0}$: with increase in $\rho_{p0}$ the decay initially slows down, reaching a minimum rate, before accelerating again. This non-monotonic variation of decay rate with $\rho_{p0}$ is quantitatively captured in the inset plot, which shows the correlation time $\tau_c$ as a function of $\rho_{p0}$. This behaviour suggests less frequent band modulation, i.e., slower dynamics, as the system approaches the phase boundary from deep within the IN-regime. In figure \ref{fig:auto_corr_fun}(b) we show the plot of $C_{acf}(t)$ \emph{vs.} $t$ for different values of $v_p$ at a fixed particle density $\rho_{p0}$. The results demonstrate that the decay of $C_{acf}(t)$ becomes progressively faster with increasing $v_p$, which is quantified by the corresponding plot of $\tau_c$ versus $v_p$ in the inset. The observed behaviour indicates that higher propulsion speed of polar particles enhances the temporal fluctuations of the bands, leading to faster dynamics and more frequent modulations, which aligns with the structural changes observed in the system: at relatively low activity of the species, the bands have a larger effective size and are less dynamic. As the activity increases, these bands become narrower, exhibiting frequent modulations.

\begin{figure}[hbt]
    \centering
    \includegraphics[width=0.8\linewidth]{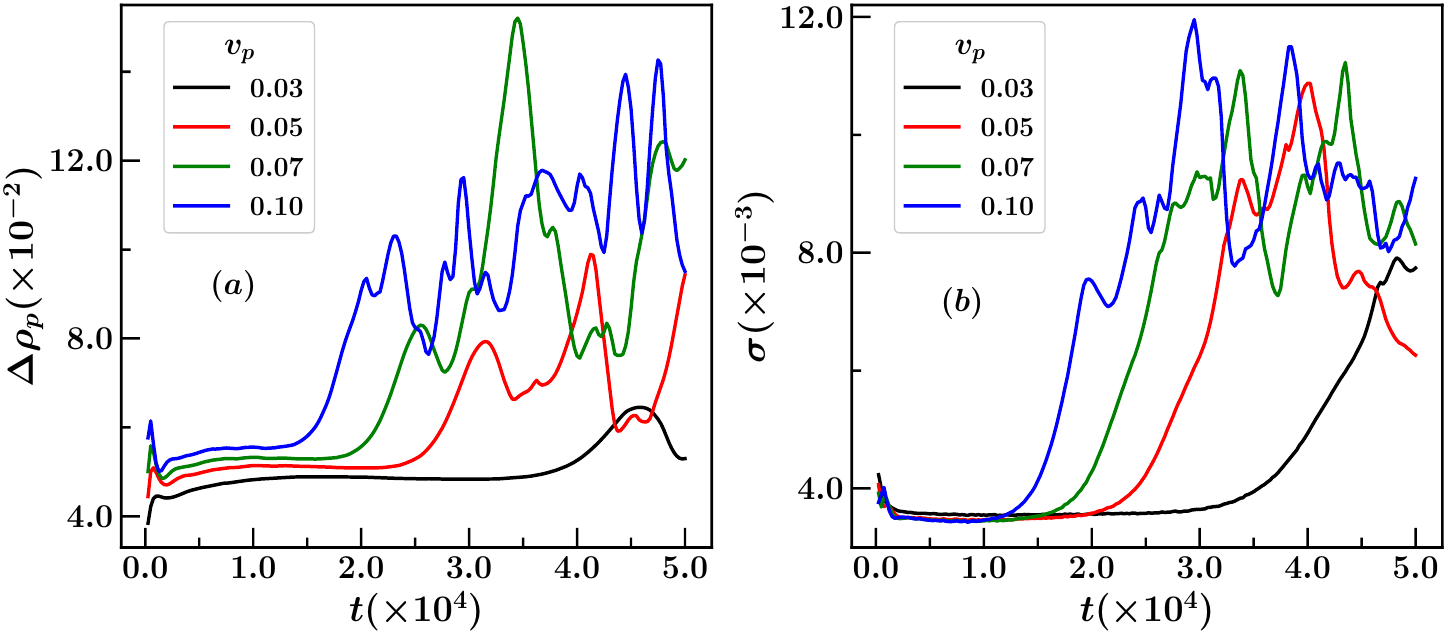}
    \caption{Variation of the mean density fluctuation of the polar species, $\Delta\rho_p$ (in panel (a)) and mean stress, $\sigma$ (in panel (b)) with time. Parameters : $L = 400$. The rest of the parameters are the same as in figure \ref{fig:pd_nop}.} 
    \label{fig:mech_time_plot}
\end{figure}

\subsubsection{\uline{Mechanism of formation of bands} :}\label{sec:mbandmec}
Before closing this section, we look into the mechanism of formation of modulating bands at high activity of polar species.\\
At relatively lower $v_p$, the steady state configuration involves a stable high-density nematically ordered band in a dilute isotropic background, where the nematic director within the band is aligned along the length of the band. To investigate the response of this configuration to increased activity of the polar species, we initiate the system (i.e. $\{\rho_p, \rho_n, \boldsymbol{P}, \boldsymbol{Q}\}$-fields) to have an initial condition that is obtained by reaching the steady-state for $v_p=0.001$. Then the value of $v_p$ is instantaneously changed to a higher value ($v_p \gtrsim 0.01$) and the system is allowed to evolve. Following the instantaneous change in $v_p$, the band initially broadens with time, subsequently developing undulations. Eventually, the band fragments into smaller, high-density nematically ordered structures that undergo frequent modulations in the low-density disordered background. The process of destabilisation of the band is presented through a series of snapshots in figure \ref{fig:band_formation}(a-e).\\
To get deeper insight into the mechanism of destabilisation of the initial nematic band, we computed the stress in apolar species, $\boldsymbol{\sigma}_{loc}(\boldsymbol{r}) = \boldsymbol{\nabla}Q_{xy}(\boldsymbol{r})$, and density fluctuation of polar species, $\delta\rho_p (\boldsymbol{r}) = \rho_p(\boldsymbol{r})-\rho_{p0}$. The evolution of the local stress field during band destabilization is presented through snapshots in figure \ref{fig:band_formation}(f-j). The corresponding global quantities are the mean stress of apolar species $\sigma = \frac{1}{L^2}\mathlarger{\sum}_{\boldsymbol{r}}\vert \boldsymbol{\sigma}_{loc}(\boldsymbol{r})\vert$, and the mean density fluctuation of polar species $\Delta\rho_p = \frac{1}{\rho_{p0}L^2}\mathlarger{\sum}_{\boldsymbol{r}}\vert\delta\rho_p(\boldsymbol{r})\vert$. The simultaneous evolution of the system configuration and the magnitude of local stress $\boldsymbol{\sigma}_{loc}$ is illustrated through an animation in the \hyperref[sm:mov3]{MOVIE-3} alongside the temporal evolution of $\sigma$ and $\Delta \rho_p$. The onset of undulations in the band is accompanied by a sharp rise in both $\sigma$ and $\Delta \rho_p$, indicating a strong coupling between band instability, mechanical stress, and density fluctuations. In figure \ref{fig:mech_time_plot}(a-b), we show the time evolution of $\Delta \rho_p$ and $\sigma$ for different values of $v_p$. As $v_p$ increases, the sharp rise in both quantities occurs earlier and becomes more pronounced, suggesting that higher activity of polar species accelerates the destabilisation of the initial band.\\
Building on these observations, we propose a mechanism for the destabilisation of the band. When the system is initiated in a well-defined banded state (at lower $v_p$), inside the band, the apolar species exhibits strong nematic ordering with the local nematic directors aligned along the length of the band. Similarly, for the polar species $\boldsymbol{P}$-field displays parallel alignment inside the band while remaining disordered outside, as shown in figure \ref{fig:pconf}. Once $v_p$ is increased to a higher value, the splay instability for polar species starts to compete with the bend instability for apolar species. It leads to the widening of the band. As the band expands, the density profile of the apolar species becomes diffuse, and the apolar directors near the band edges become increasingly susceptible to transverse fluctuations due to coupling with the disordered $\boldsymbol{P}$-field outside. In the case of a narrow, dense band, such fluctuations are suppressed by the alignment interactions with the interior apolar directors. However, as the band spreads, this stabilising effect diminishes, weakening the system’s resistance to transverse deformations. The dominance of transverse fluctuations leads to undulations in the band and the emergence of local stress gradients along the band boundaries. As a consequence, both the average stress of the apolar species and the density fluctuations of the polar species increase concurrently. These effects act cooperatively as a positive feedback mechanism, amplifying the undulations and ultimately destabilising the band.\\
To further examine the role of polar species in destabilising the band, we calculated the splay distortion energy for the polar species and the bend distortion energy for the apolar species (see Appendix \ref{app:dist_energy}). In figure \ref{fig:dis_energy}, we show the temporal evolution of the splay and bend distortion free-energy of the polar and apolar species, respectively, averaged over the entire system.  Initially, the magnitude of both energies is small; however, as time progresses, the splay contribution from the polar species begins to dominate, followed by a subsequent increase in the bend distortion of the apolar species. The corresponding spatiotemporal evolution of the two distortion-free energies is displayed in \hyperref[sm:mov4]{MOVIE-4}. The movie clearly shows the rapid increase in local splay distortion energy of polar species and in local bend distortion of apolar species at the onset of band distortion. These observations indicate that a moderate density of polar species facilitates the onset and stabilisation of the dynamical steady state.\\

\begin{figure*}[hbt]
    \centering
    \includegraphics[width=0.99\linewidth]{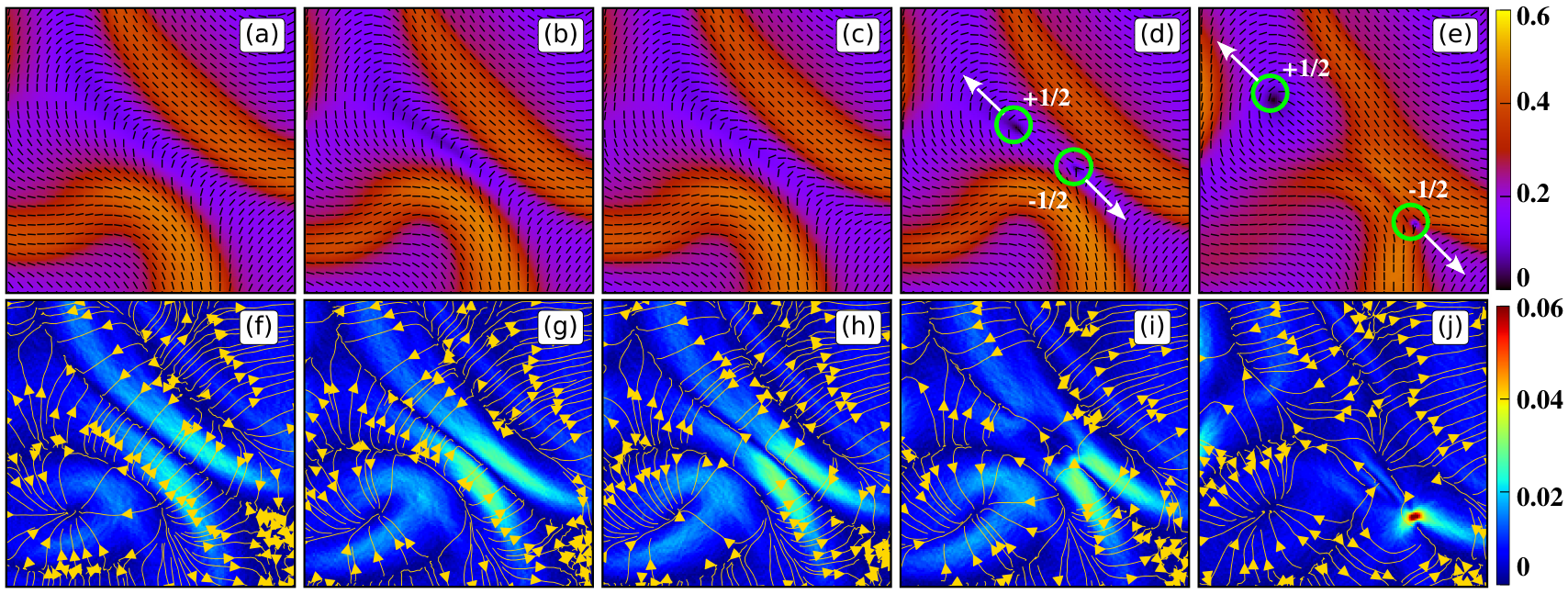}
    \caption{Visual description of the mechanism of formation of defects. The panels (a)-(e) present a zoomed-in view of the system at subsequently increasing times. The colour represents the magnitude of the local nematic order parameter ($Q$, given by Eq.\ref{eq:lnop}) according to the colour bar, and the lines represent the orientations of the local nematic director. The panels (f)-(j) depict the plot of the stress ($\boldsymbol{\nabla}Q_{xy}$) at times corresponding to the panel at the top. The colour map indicates the local stress magnitude, according to the colour bar, and the streamlines indicate the stress direction. After their generations, the $\pm \frac{1}{2}$ defects are marked, and their direction of motion is shown with arrows in panels (d-e). All snapshots correspond to a zoomed-in region of the full simulation box.}
    \label{fig:def_gen}
\end{figure*}

\begin{figure*}[hbt]
    \centering    
    \includegraphics[width=0.99\linewidth]{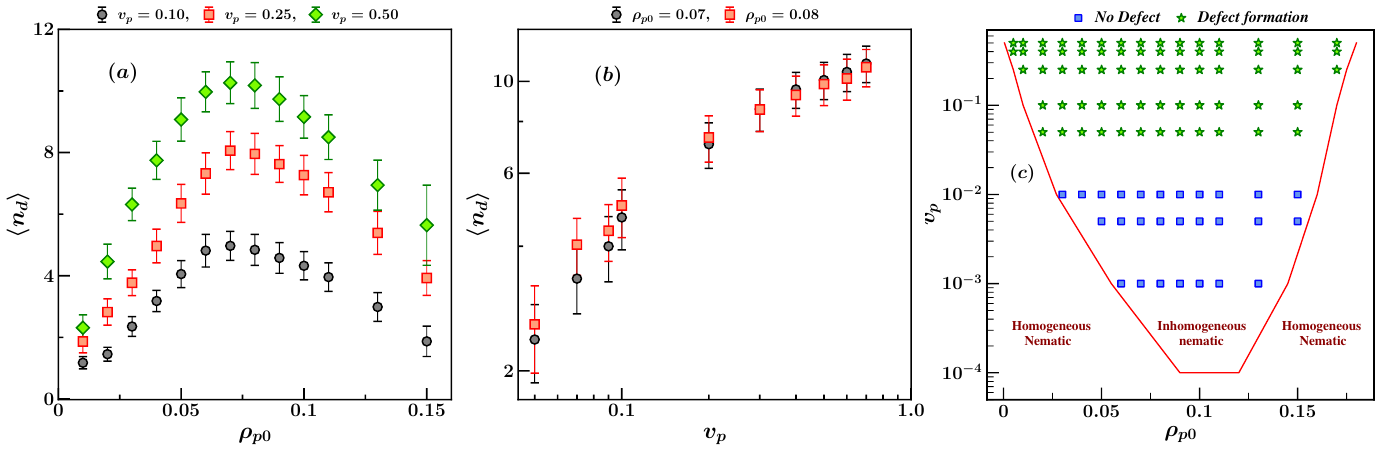}   
    \caption{Variation of mean number of $1/2$-integer topological defects in the steady state of the system, $\langle n_d \rangle$, on tuning the control parameters. Panel (a) shows the plot $\langle n_d \rangle$  \emph{vs.}  $\rho_{p0}$ for different $v_p$; Panel (b) shows the plot of $\langle n_d \rangle$\emph{vs.}$v_p$ for two different $\rho_{p0}$. The error bar represents the standard deviation calculated over time in steady state as well as independent realisations; Panel (c) presents the phase diagram of the system, highlighting the region in the $(\rho_{p0},v_p)$ space where the steady state hosts a finite number of defects. Parameters: System size, $L = 512$. The rest of the parameters are the same as in figure \ref{fig:pd_nop}.}
    \label{fig:nem_defs}
\end{figure*}

\subsection{Spontaneous proliferation of topological defects in the inhomogeneous regime}\label{sec:nemdef}
In the \emph{dynamic steady state}, we observed the spontaneous proliferation of $\pm \frac{1}{2}$ defects for $v_p \gtrsim 0.05$. We first examine the underlying mechanism responsible for the spontaneous generation of the defect pairs. Previously, we demonstrated that the presence of polar species induces bend instability in the orientation field of the apolar species \cite{mondal2024dynamical}, leading to the destabilisation of the apolar species' bands. We observe that these random modulations of the bands play a crucial role in defect proliferation. The process of proliferation of a $\pm \frac{1}{2}$ defect pair is shown through a series of snapshots figure \ref{fig:def_gen}(a–e) and the corresponding evolution of the magnitude and direction of local stress, $\boldsymbol{\sigma}_{loc}$, is shown in figure \ref{fig:def_gen}(f-j). \\
The intermediate region between the two arms of the band(s) exhibits strong bend distortion and hence generates local stress. The stress vectors of adjacent bands are directed toward each other, indicating that local stresses generate an effective attraction between neighbouring bands. Consequently, the two bands converge during modulation, increasing the intensity of the bend distortion and, hence, the elastic energy penalty and stress. The growing bend distortion leads to the formation of a wall-like structure where the distortion energy is localised. Finally, the distortion energy is released by the formation of a pair of $\pm \frac{1}{2}$ defect pairs. Immediately after the generation, the defect pairs move in opposite directions as shown in figure \ref{fig:def_gen}(d–e), leaving ordered nematic in their trail. The defects persist in the system for some time before annihilating with another defect of opposite charge. Such an annihilation event of two oppositely charged defects is displayed in figure \ref{fig:def_ann}, where the $+\frac{1}{2}$ defects approach the `Y' shaped $-\frac{1}{2}$ defect through the valley between two branches.\\
In the \emph{dynamic steady state}, the rate of proliferation and annihilation of defect pairs balance each other, maintaining a near constant number of defects in the system. These defects can be tracked using standard algorithms \cite{wenzel2021defects,delmarcelle1994topology}. The average number of defects in the steady state is calculated as, $\langle n_d \rangle = \langle  \frac{1}{2}(n_{+}(t) + n_{-}(t)) \rangle$, where $n_{\pm}(t)$ are the number of $\pm \frac{1}{2}$ defects in the system, respectively, at time $t$, and $\langle \cdots \rangle$ denotes average over time in steady state as well as multiple independent realisations. The figure \ref{fig:nem_defs} shows the variation of $\langle n_d \rangle$ with control parameters. $\langle n_d \rangle$ shows non-monotonic variation with increase in $\rho_{p0}$ and exhibits a peak as shown in figure \ref{fig:nem_defs}(a). The location of the peak is independent of $v_p$, although the peak height increases with an increase in $v_p$, which signifies the presence of a larger number of defects in the \emph{dynamic steady state} for larger $v_p$, as shown in figure \ref{fig:nem_defs}(b). The location of the peak coincides with the location of the minima in $\overline{l}_{\rho_n}$ \emph{vs.} $\rho_{p0}$ and $S$ \emph{vs.} $\rho_{p0}$. The increase of $\langle n_d \rangle$ with increase in $v_p$ can be attributed to the enhanced dynamics of the bands at higher activity levels, which increase the probability of interaction between bands and thus promote defect proliferation via the mechanism described earlier. Contrarily, within the IN-regime, no spontaneous defect proliferation is observed for $v_p \leq 0.01$. The two distinct regimes—corresponding to the presence and absence of defect proliferation—are marked in figure \ref{fig:nem_defs}(c).

\subsection{Chaotic spatiotemporal dynamic in inhomogeneous regime}\label{sec:chaos}
\begin{figure}[hbt]
    \centering
    \includegraphics[width=0.85\linewidth]{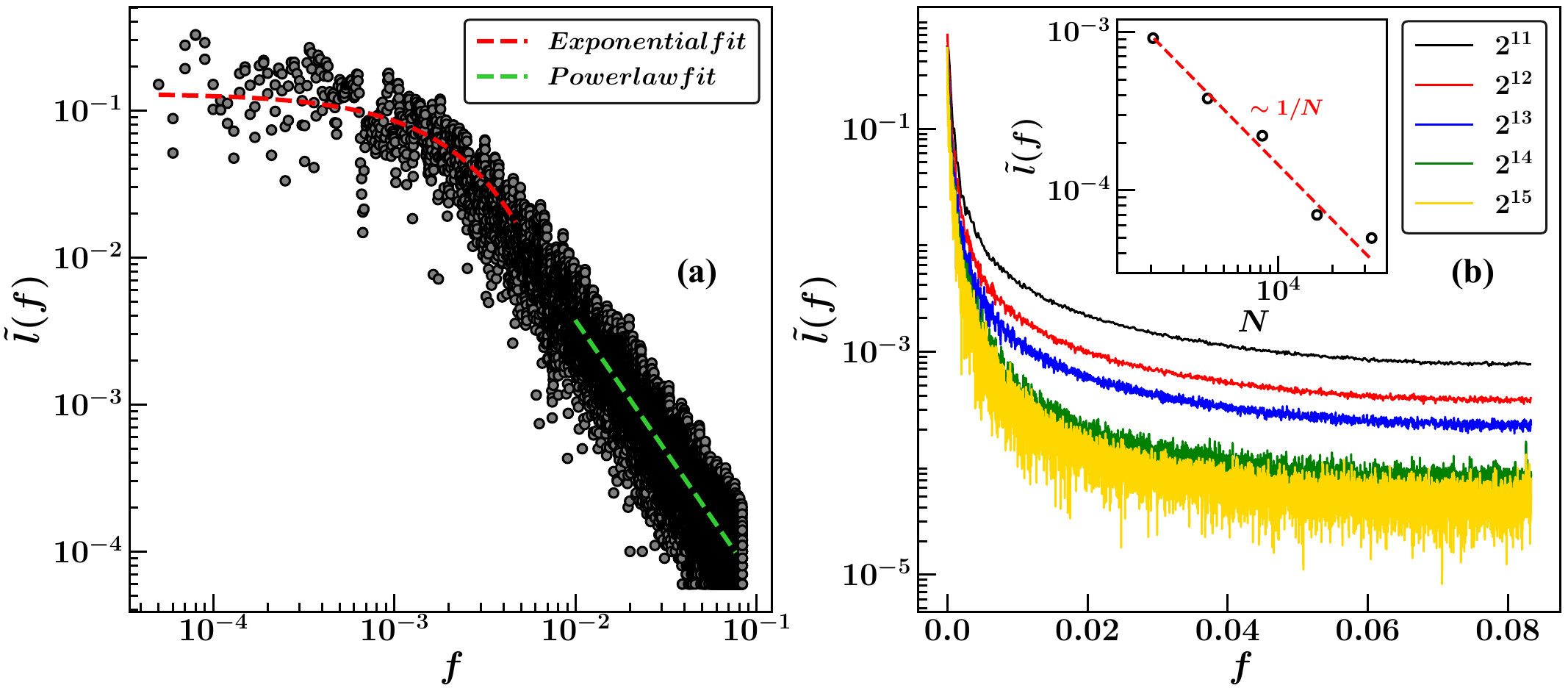}
    \caption{Characteristics of the frequency spectrum of the fluctuations in $l_{\rho_n}(t)$ in the steady state of the system. Panel (a) illustrates the piecewise stationary characteristics of the frequency spectrum. The frequency spectrum exhibits exponential characteristics at low frequencies and power-law characteristics at high frequencies; Panel (b) showcases the effect of a finite number of points in the time series on the frequency spectrum. The inset plot shows that in the tail part of the frequency spectrum $\tilde{l}(f) \sim \frac{1}{N}$. Parameters : $\rho_{p0}=0.05$, $v_p = 0.25$. The rest of the parameters are the same as in figure \ref{fig:pd_nop}.}
    \label{fig:spec_an}
\end{figure}

In this section, we explore the spatiotemporal dynamics of the system within the IN-regime. As discussed in Sec.\ref{sec:pd} $\&$ Sec.\ref{sec:inhden}, this regime is marked by the formation of high-density, nematically ordered bands that undergo frequent modulations in the \emph{dynamic steady state}. Such persistent irregular behaviour in a deterministic system is indicative of spatiotemporal chaos \cite{cross1994spatiotemporal}. For nonlinear systems with a large number of degrees of freedom, several well-established approaches exist for the characterisation of chaotic dynamics. In the present study, we employ two such approaches: the nonlinear time-series analysis method (NT)\cite{packard1980geometry,lai2003recent,skokos2016chaos} and the twin simulation method (TS)\cite{berera2018chaotic,boffetta2017chaos}. The former allows the computation of the MLE from the time series of a suitably chosen observable as well as a detailed investigation of the spectral properties of temporal fluctuations. In contrast, the latter provides a direct estimate of the maximal Lyapunov exponent (MLE) by tracking the growth of an infinitesimal perturbation introduced into the system. In what follows, we first present the spectral analysis results from the NT method, followed by the MLE results from both the NT and TS methods.\\
In the NT method, we analyse the time series of the density correlation length of the apolar species, $l_{\rho_n}(t)$. For the spectral analysis, the Fourier transform technique is used such that $\Tilde{l}(f)=\text{F.T.}\{l_{\rho_n}(t)\}$. Piecewise stationarity, a prerequisite for this analysis, is confirmed by comparing the spectra of contiguous segments of the dataset, which exhibit consistent statistical properties as shown in figure \ref{fig:pic_st}. The spectrum $\Tilde{l}(f)$ is continuous and exhibits a crossover from exponential decay at low frequencies to a power-law tail at high frequencies, as shown in figure \ref{fig:spec_an}(a). At high frequencies, $\Tilde{l}(f)$ is affected by the finite length of the dataset, scaling as $\Tilde{l}(f) \sim 1/N$ as shown in figure \ref{fig:spec_an}(b), where $N$ is the number of data points in the time series. These are well known characteristics of chaotic time series \cite{valsakumar1997signature,marwan2023power,adenij2024chaotic}.\\

\begin{figure}[hbt]
    \centering
    \includegraphics[width=0.85\linewidth]{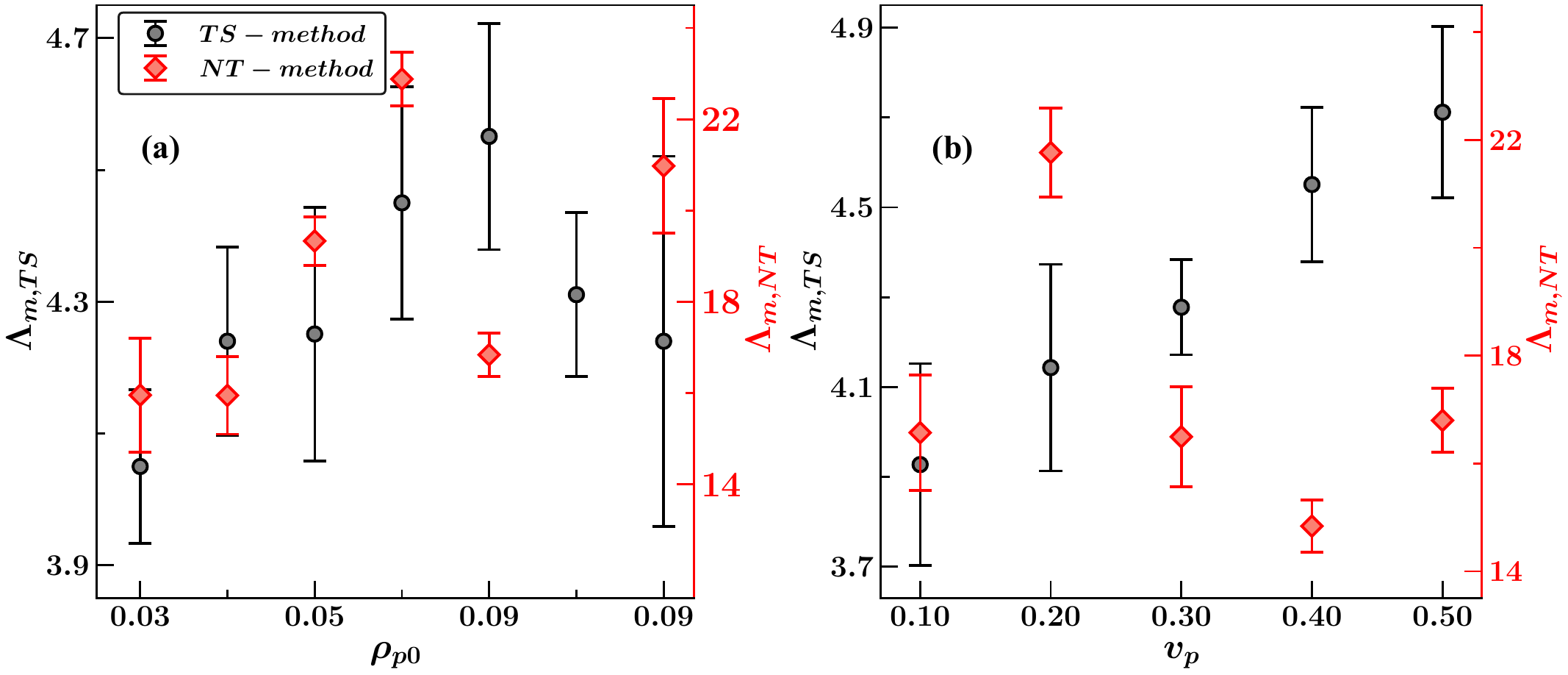}
    \caption{Variation of maximal Lyapunov exponent (MLE) on control parameters $(\rho_{p0}, v_p)$ for both $NT$ and $TS$ methods. Panel (a) shows the variation of MLE with $\rho_{p0}$ for a fixed $v_p$. Panel (b) shows the variation of MLE with $v_p$ at a fixed $\rho_{p0}$. The error bars in (a) and (b) represent the standard deviation calculated over $10$ independent realisations.}
    \label{fig:mle}
\end{figure}

To substantiate our claim of the chaotic nature of the system's dynamics, we computed the MLE using the NT and TS methods. In the NT-method, MLE is calculated by reconstructing the phase space from the time series of $l_{\rho_n}(t)$. In the reconstructed phase space, the distance between any two neighbouring points increases exponentially with time. In the TS-method, MLE is computed by adding a perturbation to the system and tracking its growth over time, which exhibits exponential behaviour in the early times. In either case, the time is rescaled by auto correlation time $\tau_c$ ( see figure \ref{fig:auto_corr_fun} and corresponding text) for the given set of parameters, i.e. $t'=\frac{t}{\tau_c}$, and the exponential growth can be expressed as $\sim \exp(\Lambda_m t')$. For NT and TS methods, the MLEs are denoted by $\Lambda_{m,NT}$ and $\Lambda_{m,TS}$, respectively. The technical details of the calculation of MLE from the NT and TS methods are presented in Appendix \ref{app:lyapunov}.\\
The dependency of $\Lambda_{m,NT}$ and $\Lambda_{m,TS}$ on the choice of control parameters $(\rho_{p0}, v_p)$  are presented in figure \ref{fig:mle}. The values of the MLE are presented as $\langle \Lambda_m \rangle \pm \Delta \Lambda_m$ where $\langle \Lambda_m \rangle$ and $\Delta \Lambda_m$ are the mean and standard deviation calculated over 10 independent realisations. Using both NT and TS methods, we obtained $\Lambda_{m,NT/TS} > 0$, thereby establishing the chaotic nature of the system's dynamics in the IN regime. Interestingly, $\Lambda_{m,TS}$ obtained from twin simulations exhibits non-monotonic behaviour with increase of $\rho_{p0}$ (see figure \ref{fig:mle}(a)), while increasing monotonically with increase of $v_p$ (see figure \ref{fig:mle}(b)). This can be understood as follows: in the \emph{dynamic steady state}, the dynamics of the bands is a dominant mechanism for the spreading and mixing of perturbations. Thus, the enhancement of structural dynamics leads to faster spreading of the perturbation and hence to a larger MLE. On the contrary, for smaller values of $v_p (\lesssim 0.02)$, the induced perturbation does not grow and $\Delta(t)$ saturates at a much smaller value.\\
On the other hand, for the NT-method, the value of $\Lambda_{m,NT}$ does not show much persistent variation with control parameters $(\rho_{p0}, v_p)$ as shown in figure \ref{fig:mle}(a-b). This qualitative difference can be understood as follows.\\
In the TS method, the MLE is computed by explicitly introducing an infinitesimal perturbation in the full dynamical system and measuring the exponential divergence of nearby trajectories during the time evolution. 
This procedure directly probes the evolution of perturbations in the solutions of the governing equations of the system, hence 
it captures the complete dynamical information of the coupled polar–nematic fields. 
Consequently, the resulting MLE captures changes in the underlying dynamics of the system with the variation of control parameters $(\rho_{p0}, v_p)$.\\
In contrast, the NT method estimates the Lyapunov exponent from a time series of a single observable of the system. This requires reconstructing an effective phase space using time-delay embedding techniques. Although such a reconstruction preserves important statistical and dynamical features of the attractor and is sufficient to detect the presence of chaotic dynamics, it represents only a projection of the full phase space. As a result, some information about the detailed structure of the underlying dynamics may not be fully retained.\\
Therefore, while the NT-based estimate reliably identifies the chaotic nature of the dynamics, it may not resolve subtle quantitative variations of the Lyapunov exponent with the control parameter as clearly as the TS method. In this sense, the TS method provides a more direct measure of the Lyapunov exponent associated with the full dynamical system, whereas the NT method yields an effective estimate based on the reconstructed dynamics.

Altogether, the comprehensive analysis in this section establishes that the dynamics in the IN-regime - particularly at higher values of $v_p$ ($\gtrsim 0.10$) - exhibit characteristics consistent with spatiotemporal chaos.

\section{Discussion}\label{sec:dis}
We investigated the spatiotemporal behaviour of a mixture of polar and apolar species using coarse-grained equations for the density and order-parameter fields, with the polar species in the minority acting as an impurity in the apolar component. The mean density and activity of the polar species are varied, while those of the apolar species are held fixed. The response of the apolar component to changes in the density of the polar species is found to be non-monotonic. In particular, within an intermediate density range of the polar species, the system attains a \emph{dynamic steady state} characterised by high-density, nematically ordered bands of the apolar species embedded in a dilute isotropic background. The bands undergo frequent modulations, and the modulation frequency increases with the activity of the polar species, leading to a defect-ridden steady state. In this regime, the dynamics of the apolar species—and of the system as a whole—appear chaotic, as supported by the spectral properties of length-scale fluctuations and the positive maximal Lyapunov exponent. \\ 

The dynamical behaviour observed in our study differs in several key aspects. Earlier reports of chaos in active systems, such as active nematic turbulence, typically occur in high-density regimes with a background fluid, where density fluctuations are often neglected and hydrodynamic interactions dominate long-range stress transmission 
\cite{shankar2022topological}. In contrast, our system operates in the dry and dilute limit, wherein the density fluctuations play a crucial role. In this setup, the chaotic state emerges through tuning the mean density and activity of polar species, highlighting the influence of interspecies coupling. Unlike active turbulence, where the length and time scale of the structure are set by both  active particles and fluid, the length and time scale of the structure of the apolar species in our system are governed primarily by the properties of the polar species, which also differentiate it from other active systems, where chaos and features of the turbulent state is controlled by the substrate properties, flow patterns and defect dynamics \cite{thijssen2021submersed,partovifard2024controlling,ronning2023defect,schimming2025turbulence,ghosh2025achieving}. \\

The observations reported in this study are of considerable importance, as they reveal a novel means of controlling the emergent behaviour of active nematic systems by introducing a secondary species. By adjusting the mean density and activity of the foreign species, one can modulate the system's dynamic states. This mechanism presents a novel approach for externally regulating active matter systems. Moreover, the predicted phenomena are amenable to experimental verification and may hold relevance in practical applications, including the development of biosensors and other soft active materials.\\
In conclusion, our study opens new avenues for future research. While our model considers a dry system without momentum conservation, it would be interesting to explore how hydrodynamic interactions influence spatiotemporal dynamics and whether the observed spatiotemporal chaos could act as a precursor to fully developed turbulence. Further, it will be interesting to study the dynamics of defects in the chaotic regime, as this can provide new insights into the system.   It will also allow a comparison of the defect dynamics in our system with those observed in turbulence states reported in other active systems. Another possible direction is to leverage the Machine Learning techniques to understand the interaction between defects  \cite{floyd2025tailoring} or dynamics of bands \cite{mcdermott2023characterizing}. In one of our future studies, we are looking forward to using Reinforcement Learning methods \cite{brunton2022data} to comprehend the defect dynamics and interaction between them.   
The power spectra examined in Fig. \ref{fig:spec_an} are interesting to explore the chaos, and in general, this could be utilised to characterize chaos in other active matter systems. The behaviour of power spectra from exponential to clean power law differentiates the dynamics in the system from chaotic to periodic, where for the latter case, one would observe a peak at a specific frequency as found in \cite{kumar2025adaptive}.

\section*{Acknowledgement}
The authors thank PARAM Shivay for the computational facility under the National Supercomputing Mission, Government of India, at the Indian Institute of Technology, Varanasi. P.S.M. thanks UGC for the research fellowship. S.M. thanks DST, SERB (INDIA), Project No.: CRG/2021/006945, MTR/2021/000438  for financial support.

\appendix
\newcommand{\appsection}[1]{%
  \refstepcounter{section}
  \section*{Appendix \thesection. #1}
  \addcontentsline{toc}{section}{Appendix \thesection. #1}
}

\appsection{Parameter Details $\&$ Methodology}\label{app:A}
The parameters in Eqs.\ref{eq:pden} and \ref{eq:pop}, governing the dynamics of polar species, are given by, $\lambda_1 = 0.50$, $\lambda_2 = 0.50$, $\lambda_3 = - 0.25$ ($= -\frac{\lambda_2}{2}$\cite{mishra2010fluctuations,marchetti2013hydrodynamics}), $D_B = 0.50$, $k_{p} = 0.50$, $D_2 = 0.50$, $\Gamma_p = 1.0$, $\beta_p = 1.0$, $\sigma_1 = 1.00$ and $D_{\rho_p} = 0.50$. The parameters in Eqs.\ref{eq:nden} and \ref{eq:nop}, governing the dynamics of apolar particles, are given by $\Gamma_Q = 1.0$, $\beta_n = 2.0$, $D_{\rho_n} = 1.0$, $D_3 = 1.0$, $a_1 = 0.90$ and $k_n = 1.00$. The activity of the apolar species is set to $a_3 = 0.60$. \\
A key aspect of the model is the choice of the coupling strength $\gamma$. For instance, in the passive limit of the model, it is reported that the evolution of the system towards the steady state exhibits a strong dependence on the strength of coupling \cite{Mishra_2025}.\newline
Minimal values of $\gamma$ result in a negligible influence of the polar species on the dynamics of the apolar species, whereas very large values lead to dominance of the apolar species over the polar species.  Our preliminary simulations revealed that for $\gamma \in [0.50,0.70]$, the steady state of the system exhibits interesting spatiotemporal characteristics. The results presented here correspond to  $\gamma = 0.65$, though the qualitative features of the system’s phase diagram in the space of control parameters remain consistent across the range mentioned above. It is worth noting that the specific range of control parameters—particularly the polar activity $v_p$ - necessary to capture the system's interesting characteristics may depend significantly on the chosen value of $\gamma$. While this dependency is a promising direction for further exploration, a detailed investigation is beyond the scope of this work. \newline 
\noindent
\uline{\emph{Test of reaching the steady state}}: To assess whether the system has attained a steady state, we analysed the temporal evolution of the probability density functions (PDFs) corresponding to the four variables: $\rho_p$, $\rho_n$, $\boldsymbol{P}$, and $\boldsymbol{Q}$. The system is considered to have reached a steady state when the PDFs of these variables exhibit no significant variation (no more than $3 - 5 \%$ variation) over time. The steady state PDFs of the apolar species are shown in figure \ref{fig:histdennop} in the Appendix \ref{app:pdf}.\\

\appsection{Calculation of Probability density function}\label{app:pdf}
The probability density function (PDF) of the observable $x\in [x_{min},x_{max}]$ is calculated by dividing the range of $x$ in $N_{bin}$ number of bins of width $w = \frac{x_{max}-x_{min}}{N_{bin}}$. The PDF of $x$ is calculated as $\mathcal{P}(x_i) = \frac{f(x_i)}{w}$, where $f(x_i)$ is the probability that the a randomly selected value of $x$ lies in the range $[x_i-\frac{w}{2},x_i+\frac{w}{2})$.\\
The PDFs of the density fluctuation and the nematic order parameter for the apolar species in different phases are shown in figure \ref{fig:histdennop}.
\setcounter{figure}{0}
\renewcommand{\thefigure}{B\arabic{figure}}
\begin{figure}[hbt]
    \centering
    \includegraphics[width=0.90\linewidth]{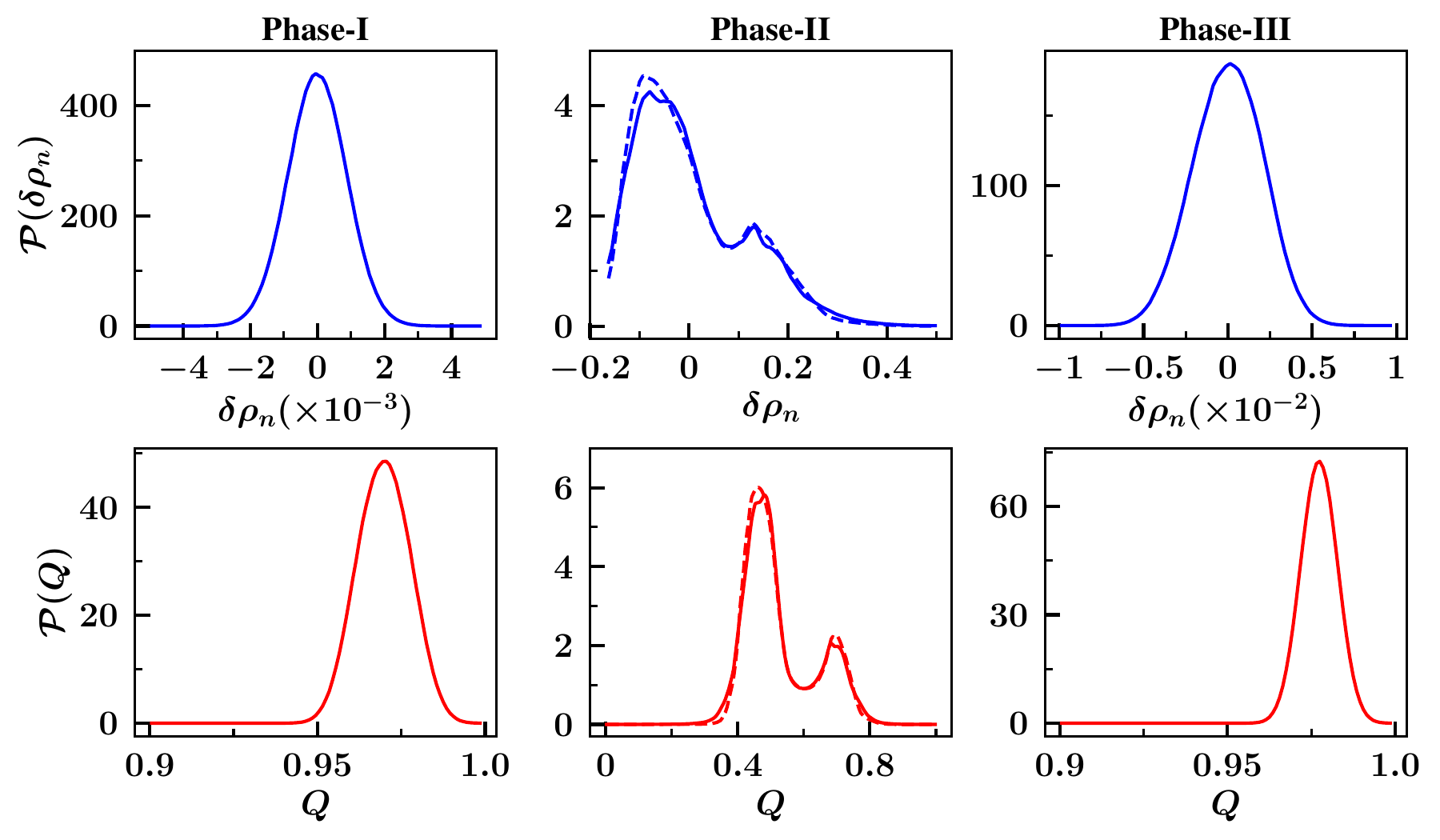}
    \caption{Probability distribution function (PDF) of nematic order parameter, $Q$, and density fluctuation, $\delta \rho_n(\boldsymbol{r}) = \rho_n(\boldsymbol{r})-\rho_{n0}$ of apolar species in different phases corresponding to snapshots shown in figure \ref{fig:pd_nop}(d-f). The solid and dashed lines are used to present data for $\Delta t = 0.10$ and $0.025$, respectively.}
    \phantomsection
    \label{fig:histdennop}
\end{figure}

\appsection{Density fluctuation in inhomogeneous regime}\label{app:denfluc}
To quantify the density inhomogeneity in the apolar species, we computed the fluctuations in their local density. For this purpose, we consider a square subcell of size $b$ centred at $(L/2, L/2)$, where $L$ is the linear dimension of the system. The density in the subcell is calculated at subsequent times to calculate $\langle\rho_n\rangle$ and $\langle\rho_n^2\rangle$, where $\langle \cdots \rangle$ denotes averaging over time in steady state. Then, the density fluctuation in the box is calculated as $\phi_{nf}= \langle\rho_n^2\rangle - \langle\rho_n\rangle^2$. The same procedure is repeated for different sizes of the subcells  $b\in [2, L/4]$. The plot of $\phi_{nf}$ \emph{vs.} $\langle\rho_n\rangle$ is shown in figure \ref{fig:num_fluc} on a log-log scale. The number fluctuation varies as $\phi_{nf} \sim \langle\rho_n\rangle^{\zeta}$, where the exponent takes the value $\zeta = 1$ for equilibrium systems. Any value of $\zeta > 1$ signifies the presence of dynamic density inhomogeneity in the system. In the IN-regime, we obtain $\zeta \approx 2$ for all values of $v_p$ as shown in the inset of figure \ref{fig:num_fluc}, which corresponds to Giant Number fluctuation(GNF) in the context of active systems \cite{ramaswamy2003active}.

\setcounter{figure}{0}
\renewcommand{\thefigure}{C\arabic{figure}}
\begin{figure}[hbt]
    \centering
    \includegraphics[width=0.59\linewidth]{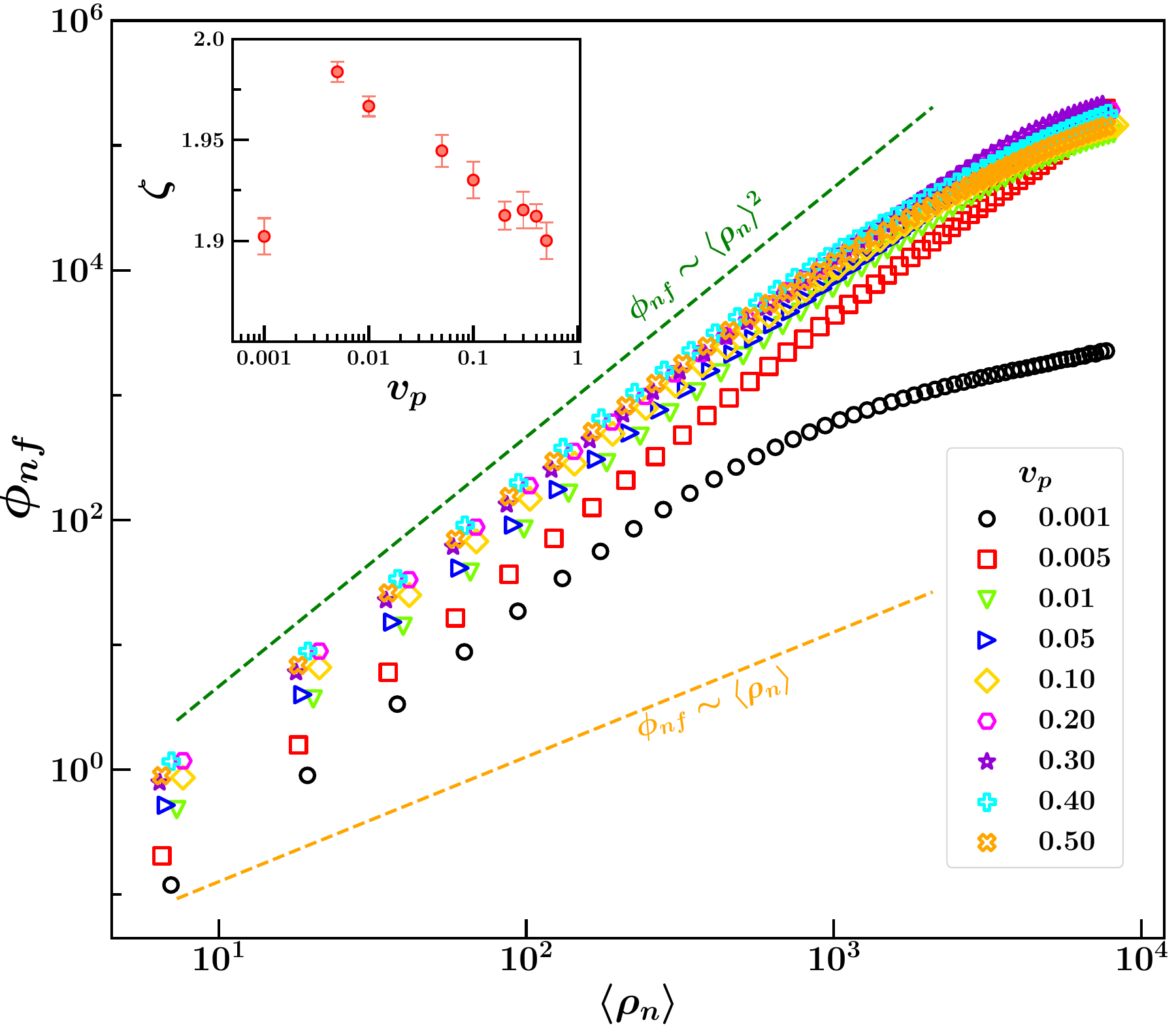}
    \caption{Plot of number fluctuation, $\sigma_{nf}$, of the apolar species in the IN-regime. The main window shows the plot of $\phi_{nf}$ \emph{vs.} $\langle \rho_n \rangle$ for different values of $v_p$ on a log-log scale. The orange and green dashed lines show the power-law fit $\phi_{nf} \sim \langle \rho_n \rangle ^{\zeta}$ with $\phi = 1$ and $2$, respectively. The inset shows the exponent $\phi$ as a function of $v_p$. Parameters : $\rho_{p0} = 0.07$, $L = 600$.} 
    \phantomsection
    \label{fig:num_fluc}
\end{figure}

\appsection{Snapshots of polar species}\label{app:psnap}
\uline{\emph{Configuration of polar species in different phases}}: In figure \ref{fig:psn_diff}, we show the snapshots of the polar species in three different phases. The heat map shows the scaled density fluctuation $\delta \rho_p/\rho_{p0}$ where $\rho_{p0}$ is the mean density of the polar species and $\delta \rho_p = \rho_p - \rho_{p0}$ is the density fluctuation of the polar species. The arrows show the direction of the local $\boldsymbol{P}$-field, with the length of the arrow proportional to the local magnitude of $\boldsymbol{P}$. The figure clearly shows that in Phase-I, the polar species is homogeneously spread across the system and remains disordered. In Phase-II, fluctuation in both density and magnitude of order parameter are prominent: inside the band, the density is higher, and the $\boldsymbol{P}$-field shows significant ordering, whereas outside the band, the density is low and $\vert \boldsymbol{P} \vert \approx 0$. In Phase-III, the homogeneous configuration of the polar species is restored, but in contrast to Phase-I, the $\boldsymbol{P}$-field shows strong ordering.

\FloatBarrier
\setcounter{figure}{0}
\renewcommand{\thefigure}{D\arabic{figure}}
\begin{figure}[hbt]
    \centering
    \includegraphics[width=0.99\linewidth]{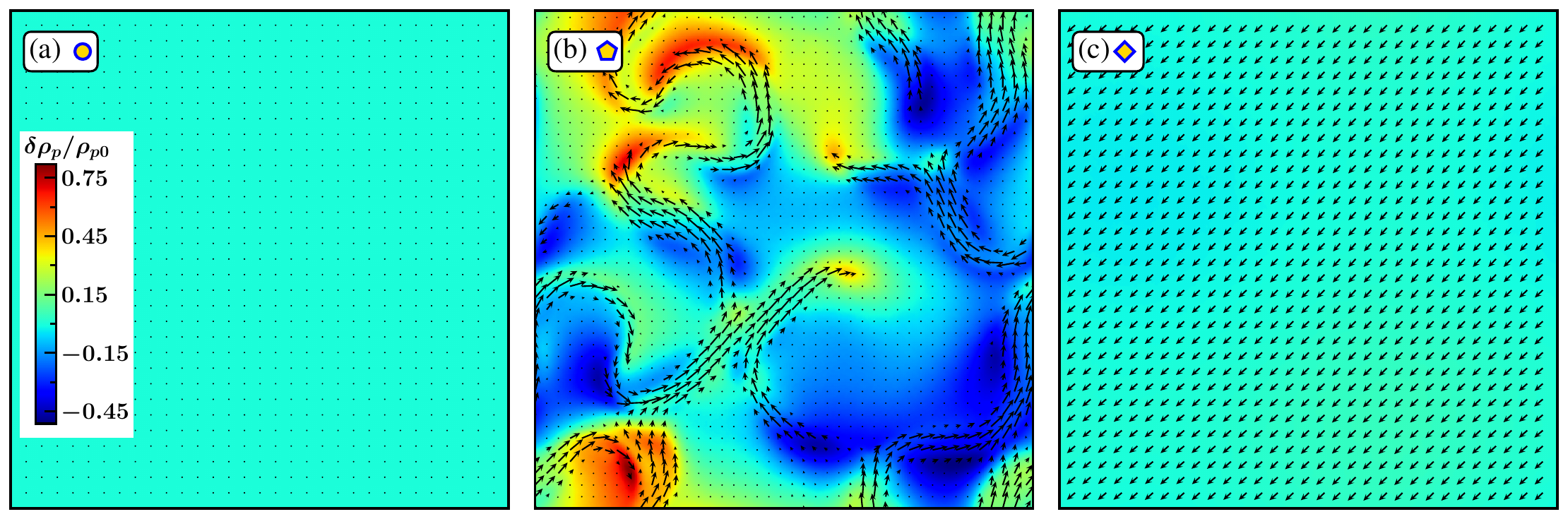}
    \caption{Snapshots of the polar species in different phases. The three panels (left to right) correspond to the same set of parameters as in figure \ref{fig:pd_nop}(d-f), respectively. The heat map shows the normalised density fluctuation of polar species $\frac{\delta \rho_p}{\rho_{p0}}$. The arrows represent the direction of the local $\boldsymbol{P}$-field, and the length of the arrows is proportional to the magnitude of the local $\boldsymbol{P}$-field. The snapshots capture the entire simulation domain.} 
    \phantomsection
    \label{fig:psn_diff}
\end{figure}
\FloatBarrier

\begin{figure}[hbt]
    \centering
    \includegraphics[width=0.99\linewidth]{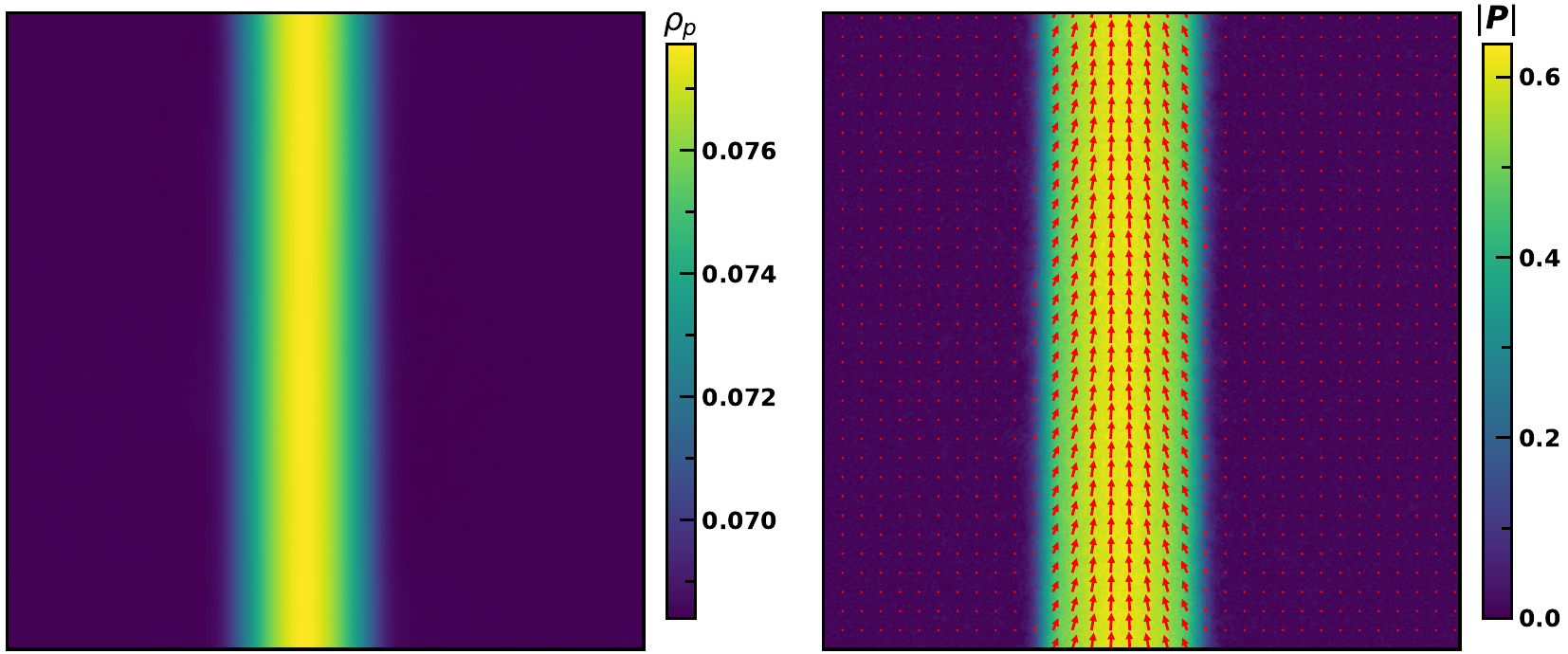}
    \caption{Configuration of the polar species for the artificial initial condition discussed in section \ref{sec:mbandmec}. (left panel) The heat map shows the local density, $\rho_p(\boldsymbol{r})$, of the polar species as indicated by the colorbar. (right panel) The heat map shows the local magnitude of the polarisation field, $\vert\boldsymbol{P}\vert$, as indicated by the colorbar. The arrows show the direction of the local $\boldsymbol{P}$ and the length of the arrows is proportional to the local magnitude of the $\boldsymbol{P}$-field.. The parameters are the same as in figure \ref{fig:band_formation}. The snapshots display the complete simulation box.} 
    \phantomsection
    \label{fig:pconf}
\end{figure}

\uline{\emph{Configuration of polar species for artificial initial condition}}: In figure \ref{fig:pconf}(a-b) we show the snapshot of the initial condition of the polar species for the artificial initial condition as discussed in section \ref{sec:mbandmec} in the main text where the panel (a) shows configuration of enematic density, and panel (b) shows configuration of $\boldsymbol{P}$-field. 

\appsection{Distortion energy}\label{app:dist_energy}
The configuration of the orientation of the $\boldsymbol{P}$-field for the splay instability in polar species and the nematic director field for bend instability in apolar species is shown in figure \ref{fig:dis_fig}.

\setcounter{figure}{0}
\renewcommand{\thefigure}{E\arabic{figure}}
\begin{figure}[hbt]
    \centering
    \includegraphics[width=0.90\linewidth]{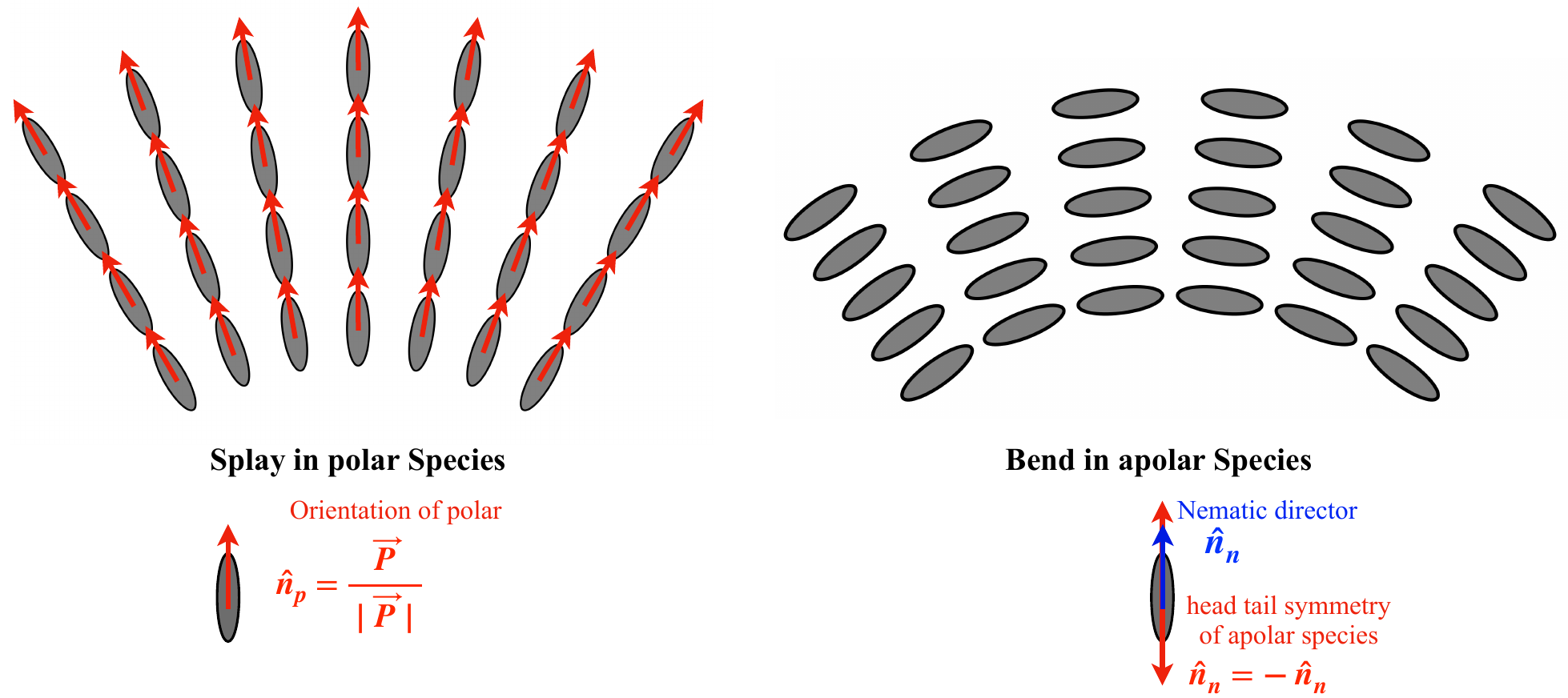}
    \caption{Schematic diagram showing the configuration of local orientation of polar species, $\hat{n}_p$, for splay distortion and local nematic director field for apolar species.} 
    \phantomsection
    \label{fig:dis_fig}
\end{figure}

\begin{figure}[hbt]
    \centering
    \includegraphics[width=0.50\linewidth]{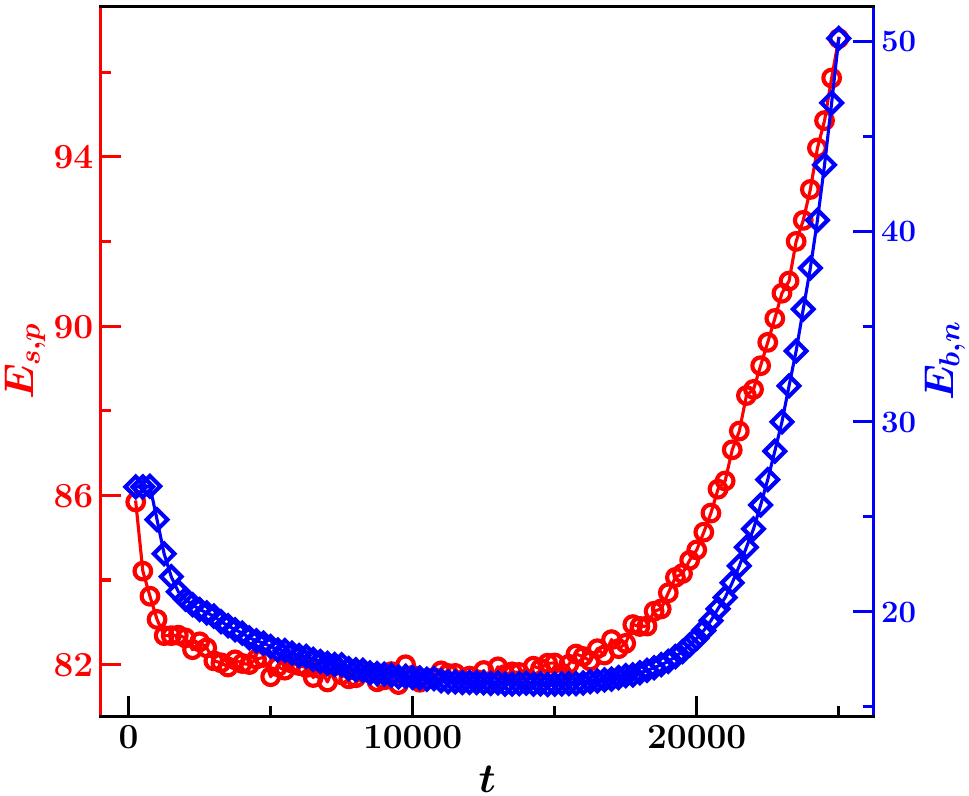}
    \caption{Time evolution of total splay distortion energy in polar species, $E_{s,p}$, and total bend distortion energy in apolar species, $E_{b,n}$. Parameters: $\rho_{p0}=0.070$,$v_p=0.070$, $L=400$.  }
    \phantomsection
    \label{fig:dis_energy}
\end{figure}

The local splay distortion energy for polar species is defined as $\xi_{s,p}(\boldsymbol{r})=(\boldsymbol{\nabla}\cdot\bar{\boldsymbol{P}})^2$, where $\bar{\boldsymbol{P}}(\boldsymbol{r})=\rho_p(\boldsymbol{r})\boldsymbol{P}(\boldsymbol{r})$. For the apolar species, local bend distortion energy for apolar species is give byn $\xi_{b,n}(\boldsymbol{r})=\mathlarger{\sum}_{\alpha,\beta,\gamma}[(\partial_{\alpha}\bar{Q}_{\alpha \beta})(\partial_{\gamma}\bar{Q}_{\gamma \beta}) + (\partial_{\beta}\bar{Q}_{\alpha \gamma})(\partial_{\beta}\bar{Q}_{\alpha \gamma})]$, where $\bar{\boldsymbol{Q}}(\boldsymbol{r})=\rho_n(\boldsymbol{r})\boldsymbol{Q}(\boldsymbol{r})$.\newline 
\noindent
The total splay distortion energy for polar species is computed as $E_{s,p}=\frac{1}{L^2}\mathlarger{\sum}_{\boldsymbol{r}}\xi_{s,p}(\boldsymbol{r})$. Similarly, total bend distortion energy for the apolar species is calculated as, $E_{b,n}=\frac{1}{L^2}\mathlarger{\sum}_{\boldsymbol{r}}\xi_{b,n}(\boldsymbol{r})$.\newline
figure \ref{fig:dis_energy} shows the time evolution of $E_{s,p}$ and $E_{b,n}$, starting with the artificial initial condition as discussed in Sec.\ref{sec:inhden}.

\appsection{Defect annihilation}\label{app:def_ann}
figure \ref{fig:def_ann}(a-e) show the annihilation event of a pair of $\pm \frac{1}{2}$ defects in the mixture. Before an annihilation event, the $+1/2$ defect approaches the `Y'-shaped $-1/2$ defect through the valley between the two branches of the latter.

\setcounter{figure}{0}
\renewcommand{\thefigure}{F\arabic{figure}}
\begin{figure}[hbt]
    \centering
    \includegraphics[width=0.99\linewidth]{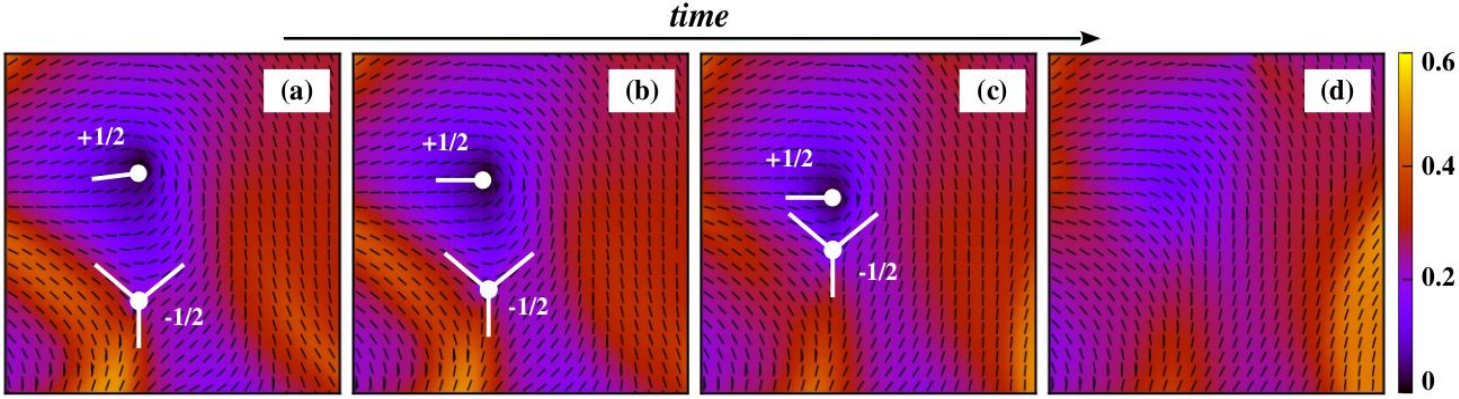}
    \caption{Series of snapshots showing annihilation event of a pair of $\pm \frac{1}{2}$ defects. plots (a-e) show the snapshots at successive intervals. The snapshots are zoomed in on the region close to the defect pair. The heatmap shows the magnitude of the nematic order parameter according to the colorbar, and the lines show the direction of the local nematic director.}
    \phantomsection
    \label{fig:def_ann}
\end{figure}

\appsection{Piecewise stationarity of frequency spectrum}\label{app:piece_stat}
A prerequisite for applying the Fourier transformation to a time series is that the resulting frequency spectrum exhibits piecewise stationarity. This condition ensures that the statistical features of the frequency spectrum remain invariant with respect to the length of the time series. The piecewise stationarity of the frequency spectrum obtained from the time series of the nematic density correlation length $l_{\rho}(t)$ is shown in figure \ref{fig:pic_st}. The $l(t)$ \emph{vs.} $t$ plot shown in the main frame is divided into four segments of equal length, highlighted with different colours, and the frequency spectrum, $l_{\rho_n}(f)$, for each segment is shown in the corresponding inset. The similarity in the spectral characteristics across different segments confirms the piecewise stationarity of the time series.

\setcounter{figure}{0}
\renewcommand{\thefigure}{G\arabic{figure}}
\begin{figure}[hbt]
    \centering
    \includegraphics[width=0.85\linewidth]{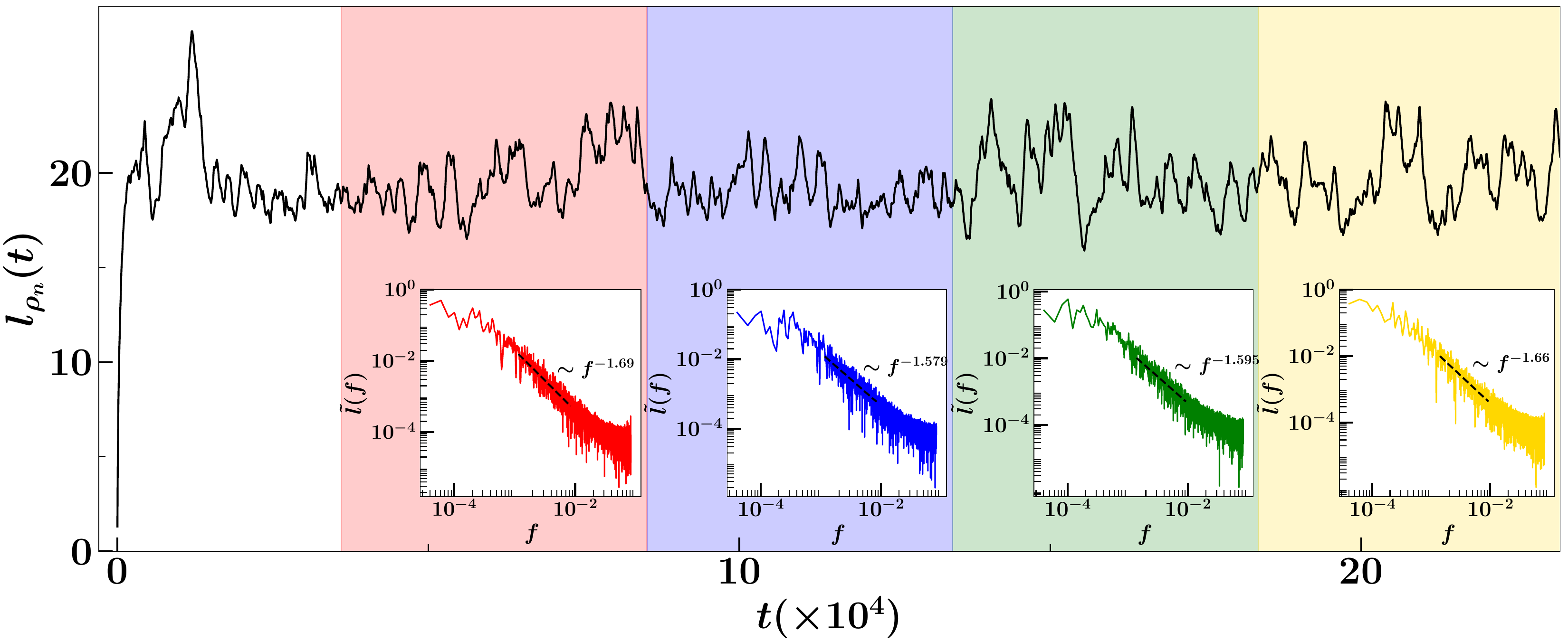}
    \caption{The piecewise stationarity of the frequency spectrum of fluctuations in the steady state length scale $l_{\rho_n}(t)$. The $l_{\rho_n}(t)$ \emph{vs.} $t$ plot is shown in the main frame. The frequency spectrum for regions highlighted with different colours is shown in the corresponding inset.}
    \phantomsection
    \label{fig:pic_st}
\end{figure}

\appsection{Calculation of maximal Lyapunov exponent}\label{app:lyapunov}
Here, we present the methodology used to calculate the maximal Lyapunov exponent.\newline
\noindent
\uline{\emph{Twin simulation method}}: In the steady state, the instantaneous configuration of the system can be described as $\boldsymbol{X}=\{\rho_{n,i}, \rho_{p,i}, \boldsymbol{P}_i, \boldsymbol{Q}_i, i=1,\cdots, N \}$, where $N$ is the number of lattice points.\\
We create a perturbed copy of the system, $\bar{\boldsymbol{X}}=\boldsymbol{X}+\delta\boldsymbol{X}$, where $\delta\boldsymbol{X}$ denoted the perturbation. The perturbation is introduced by temporarily switching off the activity of the polar species for a short duration $\delta t$ such that $\delta t \ll\tau_c$, where $\tau_c$ is the correlation time for the corresponding set of parameters (see figure \ref{fig:auto_corr_fun} and corresponding text).\\
With time, $t$, the perturbation grows, which can be tracked by calculating the decorrelator defined as $\Delta (t) = \langle \vert \boldsymbol{X}-\bar{\boldsymbol{X}} \vert^2 \rangle$, where $\langle \cdots \rangle$ denotes averaging over all the lattice points as well as independent realisations. At early times $\Delta(t)$ grows exponentially $\Delta(t) \sim \Delta(0)\exp(\Lambda_{m,TS}t')$, where $\Lambda_{m,TS}$ is the maximal Lyapunov exponent and $t'=\frac{t}{\tau_c}$ is the rescaled time. However, at late times $\Delta(t)$ saturates when the two copies ($X$ and $X'$) become completely decorrelated. The variation of $\Delta(t)$ with time, $t$, is shown in figure \ref{fig:ts_dec}(a) for $v_p = 0.10$ and $\rho_{p0}=0.07$, wherein the exponential growth regime is clearly visible. $\Lambda_{m,TS}>0$ indicates the chaotic characteristics of the systems dynamics.\\

\setcounter{figure}{0}
\renewcommand{\thefigure}{H\arabic{figure}}
\begin{figure}[hbt]
    \centering
    \includegraphics[width=0.850\linewidth]{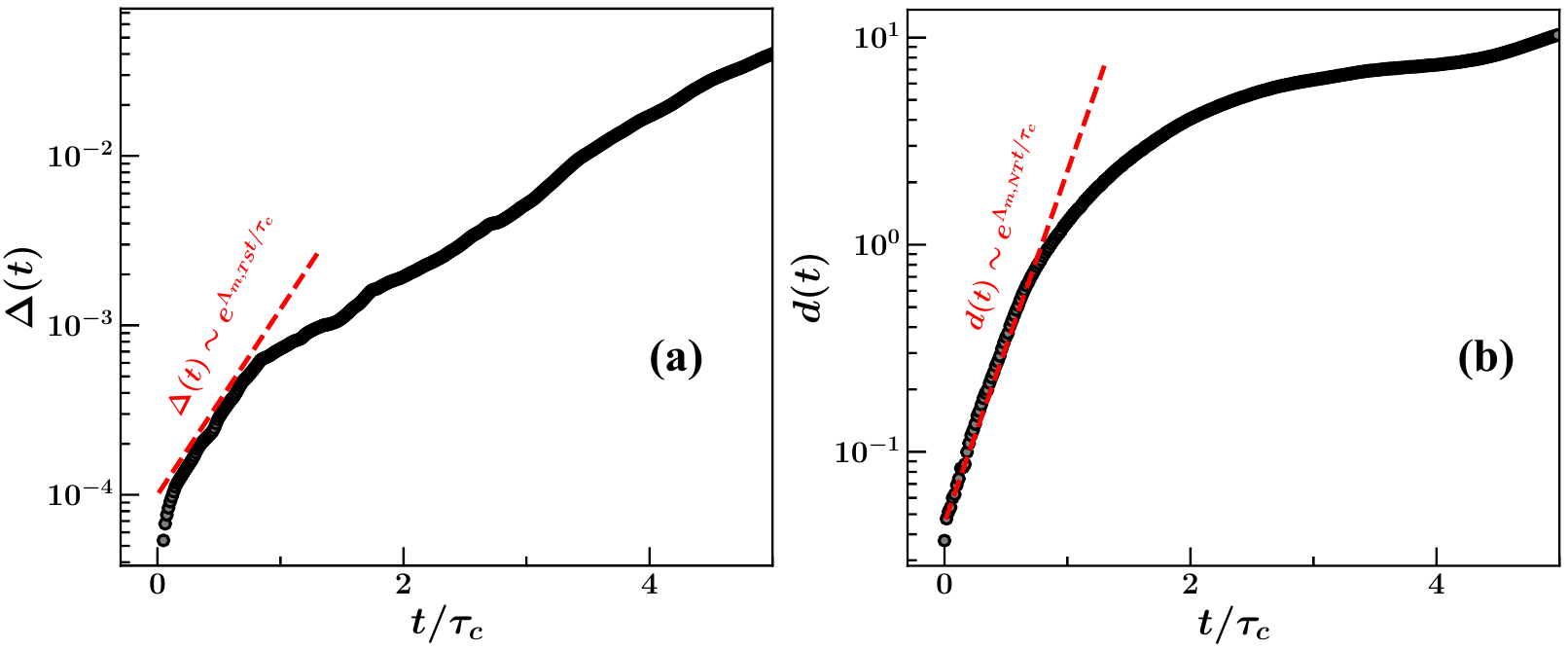}
    \caption{Calculation of MLE by NT and TS methods. Panel (a) shows the exponential growth of decorrelator $\Delta(t)$ with rescaled time $t'=\frac{t}{\tau_c}$ as $\Delta(t) \sim \Delta(0) e^{\Lambda_{m,TS}t/{\tau_c}}$ on a log-y scale. The red dashed line shows the exponential fit. Panel (b) shows the exponential growth of distance $d(t)$ with rescaled time $t'=\frac{t}{\tau_c}$ as $d(t) \sim d(0) e^{\Lambda_{m,NT}t/{\tau_c}}$ on a log-y scale. The red dashed line shows the exponential fit.}
    \phantomsection
    \label{fig:ts_dec}
\end{figure}

\begin{figure}[hbt]
    \centering
    \includegraphics[width=0.90\linewidth]{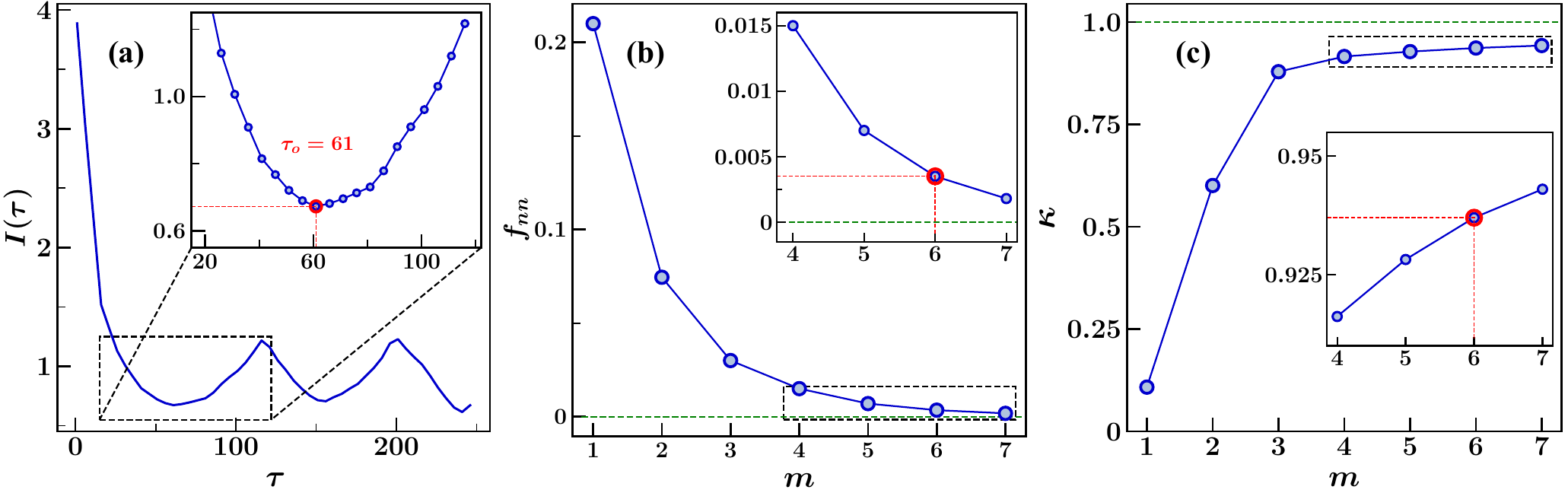}
    \caption{Results of the nonlinear time series analysis. Panel (a) illustrates the determination of the optimal time delay $\tau$ from the mutual information $I(\tau)$. Panel (b) shows the estimation of the optimal embedding dimension $m$ based on the decay of the fraction of false nearest neighbours $f_{nn}$. Panel (c) presents the determinism coefficient $\kappa$, indicating that for the chosen values of $m$ and $\tau$, $\kappa \approx 0.92$.}
    \phantomsection
    \label{fig:nta}
\end{figure}

\noindent
\uline{\emph{Nonlinear time series analysis}}: The calculation of the Lyapunov exponent from the time series of an observable involves the following steps - (i) reconstructing the phase space and determining its optimal embedding dimension, (ii) performing a determinism test, and (iii) estimating the Lyapunov exponent.\\
Phase-space reconstruction is carried out using Taken’s \cite{kantz2003nonlinear,casdagli1992nonlinear}. Steps (i) and (ii) are critical for ensuring a successful reconstruction, as they yield the appropriate time delay $\tau$ and embedding dimension $m$. The optimal value of $\tau$ is chosen as the first minimum of the mutual information $I(\tau)$ \cite{fraser1986independent}, while the optimal embedding dimension $m$ is obtained from the decay of the fraction of false nearest neighbours, denoted by $f_{nn}$, with increasing dimension of the reconstructed phase space \cite{kennel1992determining}. Once the phase space is reconstructed, a determinism test is performed to verify whether the time series originates from an underlying deterministic system or is dominated by stochastic fluctuations \cite{kaplan1992direct}.\\
The calculation of $\tau$ and $m$ and the results of the determinism test for our system are presented in figure \ref{fig:nta} for one set of parameters. Across the range of control parameters $(\rho_{p0}, v_p)$, the optimal embedding dimension was consistently found to be $m = 6$, while the time delay $\tau$ lies within the interval $\tau \in[41, 65]$. The determinism test, conducted using these embedding parameters, yields determinism coefficients $\kappa$ in the range $[0.92, 0.94]$ across different parameter sets, thereby confirming the deterministic character of the dynamics.

Finally, the Lyapunov exponent is obtained by monitoring the divergence of pairs of trajectories originating from two neighbouring points in the reconstructed phase space. The separation between these trajectories grows exponentially as as $d(t)=d(0)\exp(\Lambda_{m,NT}t')$ as shown in figure \ref{fig:ts_dec}(b), where, $\Lambda_{m,NT}$ denotes the maximal Lyapunov exponent and and $t'=\frac{t}{\tau_c}$ is the rescaled time. A positive value of $\Lambda_{m,NT}$ signifies the chaotic character of the underlying dynamics.\\

\appsection{Description of Supplementary Movies}\label{app:suppanim}
\textbf{\uline{MOVIE-1}\label{sm:mov1} : Spatial structures in the system for different activity}. The supplementary movie shows the animation of the nematic order parameter $Q(\boldsymbol{r})$ field for different activities of polar species, $v_p$. In the animation for $v_p=0.05$, $0.20$, and $0.50$ the $+\frac{1}{2}$ and $-\frac{1}{2}$ defects are marked with `\tikz\draw[fill=red, draw=black, line width=1.2pt] (0,0) circle (0.12cm);' and `\tikz\draw[fill=yellow, draw=black, line width=1.2pt] 
(0,0) -- (0.24,0) -- (0.12,0.21) -- cycle;' symbols, respectively.\newline
\uline{\emph{Parameters}} : System Size, $L=400$. Rest of the parameters are same as in figure \ref{fig:pd_nop}.\newline
\uline{\emph{Link}} :
\url{https://drive.google.com/file/d/14aPSW9hEPjI7D4qH7duYDDqFml6zJ49I/view?usp=sharing}

\vspace{1.5em}

\noindent
\textbf{\uline{MOVIE-2}\label{sm:mov2} : Emergence of the structures in the system}. The supplementary movie illustrates the evolution of the system from a homogeneous isotropic initial condition toward the \emph{dynamic steady state}, as well as the temporal development of the characteristic length scales of the system. The left and middle panels display the evolution of the density field, $\rho_n(\boldsymbol{r})$, and the nematic order parameter field, $Q(\boldsymbol{r})$, respectively, of the apolar species. The right panel shows the time evolution of the correlation lengths of the density, $l_{\rho_n}(t)$, and the nematic order parameter, $l_Q(t)$.\newline
\uline{\em Parameters} : System Size, $L=512$. Rest of the parameters are same as in figure \ref{fig:pd_nop}.\newline
\uline{\em Link} :
\url{https://drive.google.com/file/d/1CYK3jdPHYQP1tJ_gQFj8Zk2UaUa1gUV7/view?usp=sharing}

\vspace{1.5em}

\noindent
\textbf{\uline{MOVIE-3}\label{sm:mov3} : Destabilization of band in the inhomogeneous regime}. The supplementary movie depicts the evolution of the system toward a dynamic steady state, starting from an initial condition corresponding to the steady-state configuration at a low self-propulsion speed, $v_p = 0.001$. Following a rapid increase of polar activity to $v_p = 0.07$, the system evolves into a \emph{dynamic steady state}. The left panel presents the temporal evolution of the nematic order parameter field, $Q(\boldsymbol{r})$. In contrast, the middle panel shows the evolution of the stress field $\boldsymbol{\nabla}Q_{xy}$. The right panel displays the time evolution of the mean stress, $\sigma$, and the mean density fluctuation of the polar species, $\Delta \rho_p$.\newline
\uline{\em Parameters} : System Size, $L=400$. Rest of the parameters are same as in figure \ref{fig:pd_nop}.\newline
\uline{\em Link} :\newline
\url{https://drive.google.com/file/d/1rH0AjsFjAHiDWaqoz0x7HJneRWWZcwPY/view?usp=sharing}

\vspace{1.5em}

\noindent
\textbf{\uline{MOVIE-4}\label{sm:mov4} : Distortion energy in polar and apolar species}. The supplementary movie shows the evolution of energy corresponding to splay distortion in $\bar{\boldsymbol{P}}$ for the polar species and bend distortion in $\bar{\boldsymbol{Q}}$ for apolar species. The mathematical expressions for the distortion energies are given in Appendix \ref{app:dist_energy}.\newline
\uline{\em Parameters}: Mean density of polar species $\rho_{p0}=0.07$, activity of polar species $v_p=0.07$, system size $L=400$. The details of the rest of the parameters are the same as in figure \ref{fig:band_formation}.\newline
\uline{\em Link} :\newline
\url{https://drive.google.com/file/d/1XJ0tTafVTynRojecxsclBoS2pa9P2e0H/view?usp=sharing}

\bibliographystyle{iopart-num}
\bibliography{References}

\end{document}